\documentclass[english,showpacs,onecolumn,superscriptaddress]{revtex4-2}

\usepackage[english]{babel}

\usepackage{amsfonts}
\usepackage{verbatim}
\usepackage{amsmath}
\usepackage{amsthm}
\usepackage{amssymb}
\usepackage{graphicx}
\usepackage{hyperref}
\usepackage{subfigure}
\usepackage{cancel}
\usepackage[usenames,dvipsnames]{color}

\renewcommand{\phi}{\varphi}

\usepackage{natbib}

\begin{document}
\title{The double-faced electrostatic behavior of PNIPAm microgels.}

\author{Simona Sennato}
\affiliation{CNR-ISC Sede Sapienza and Dipartimento di Fisica, Sapienza Universit\`{a} di Roma,
P.le A. Moro 2, 00185 Roma, Italy}
\author{Edouard Chauveau}
\affiliation{Laboratoire Charles Coulomb (L2C), UMR 5221 CNRS-Universit\'{e} de Montpellier,
Montpellier, France}
\author{Stefano Casciardi}
\affiliation{National Institute for Insurance against Accidents at Work (INAIL), Department of Occupational and Environmental Medicine, Epidemiology and Hygiene, 00078 Monteporzio Catone (Roma), Italy}
\author{Federico Bordi}
\affiliation{CNR-ISC Sede Sapienza and Dipartimento di Fisica, Sapienza Universit\`{a} di Roma,
P.le A. Moro 2, 00185 Roma, Italy}
\author{Domenico Truzzolillo}
\email{domenico.truzzolillo@umontpellier.fr}
\affiliation{Laboratoire Charles Coulomb (L2C), UMR 5221 CNRS-Universit\'{e} de Montpellier,
Montpellier, France}


\date{\today}

\begin{abstract}
PNIPAm microgels synthesized via free radical polymerization (FRP) are often considered as neutral colloids in aqueous media, although it is well known, since the pioneering work of Pelton and coworkers [Langmuir \textbf{1989}, 5, 816-818],
that the vanishing electrophoretic mobility characterizing swollen microgels largely increases above the lower critical solution temperature (LCST) of PNIPAm, at which microgels partially collapse. The presence of an electric charge has been attributed to the ionic initiators that are employed when FRP is performed in water and that stay anchored to microgel particles.
Combining dynamic light scattering (DLS), electrophoresis, transmission electron microscopy (TEM) and atomic force microscopy (AFM) experiments, we show that collapsed ionic PNIPAm microgels undergo large mobility reversal and reentrant condensation when they are co-suspended with oppositely charged polyelectrolytes (PE) or nanoparticles (NP), while their stability remains unaffected by PE or NP addition at lower temperatures, where microgels are swollen and their charge density is low. Our results highlight a somehow double-faced electrostatic behavior of PNIPAm microgels due to their tunable charge density: they behave as quasi-neutral colloids at temperature below LCST, while they strongly interact with oppositely charged species when they are in their collapsed state. 
The very similar phenomenology encountered when microgels are surrounded by polylysine chains and silica nanoparticles points to the general character of this twofold behavior of PNIPAm-based colloids in water.
\end{abstract}

\maketitle

\section{Introduction}
Poly(N-isopropylacrylamide)(PNIPAm) microgels synthesized via free radical polymerization (FRP) are thermoresponsive colloids used nowadays for many research purposes and in different areas of investigation, ranging from nonlinear optics and soft matter to nanomedicine and materials science~\cite{fernandez-nieves_microgel_2011,pich_chemical_2011,guan_pnipam_2011}. Microgels are miniature hydrogels with a size ranging from tens of nanometres to several microns~\cite{saunders_microgel_1999,karg_new_2009}. Like bulky gels, microgels are usually biocompatible~\cite{weng_tissue_2004}, however, due to their small size, they exhibit many advantages over macroscopic gels when used as biomaterials. One major advantage is that microgel response to external stimuli is much faster than that of bulky gels~\cite{wang_temperature-jump_2001,reese_nanogel_2004}, since the rate of volume change scales as $l_g^2$, with $l_g$ being the relevant length scale of the whole polymer network~\cite{tanaka_kinetics_1979}.


{In the context of microgel synthesis, stimuli-responsive polymers have attracted much attention over the last decades due to a wide range of innovative applications in diagnostics, tissue engineering, and 4D-printed soft devices. Their physical properties, such as shape, size, and hydrophobicity, can be changed using external stimuli, including thermal, electrical, magnetic, and ionic interactions~\cite{prabaharan2006a,motornov2003a,szabo1998a,siegel1988a,sutani2001a}. So far, various applications have been demonstrated for micromachines~\cite{guan2012a,ter2015a}, structures of biomedical interest~\cite{kirillova2017a}, inchworm-like motion~\cite{yamada2009a}, and soft actuator devices~\cite{guo2018a,hu2019a,yang2016a,yamada2008a}.}


In the past decade, the synthesis of microgels with complex architectures received considerable attention. Among all the different known microgel morphologies, core/ shell~\cite{richtering_nanoparticles_2020,crassous_anisotropic_2015} or multishell~\cite{cors_determination_2018,lima_updated_2020} structures represented the core of this line of research.

{Despite that, the synthesis of pure PNIPAm microgels via FRP stays one of the most used methods to produce soft spherical colloids for research purposes, since they offer the lowest level of structural complexity, the smallest size ratio between the collapsed and the swollen state, a relatively weak dependence of their size on pH and ionic strength, and a well-understood inter-particle potential, collective dynamics in concentrated suspensions and 2D organization at the air-water interface~\cite{bergman_new_2018,philippe_glass_2018,picard_organization_2017}.
Most of these studies focused on the swelling properties of PNIPAm microgels and how the latter can trigger large variation of optical or rheological properties of materials, while their electrostatic properties have been left relatively unexplored.
Though since the pioneering work of Pelton and co-workers~\cite{pelton_particle_1989}, a large increase of microgel electrophoretic mobility is known to exist above the volume phase transition temperature of the microgels due to the ionic initiators anchored to the polymer network, PNIPAm microgels have been often considered as neutral colloids, since their inner structure does not contain ionizable monomers. Only very recently few works pointed out that the residual charge borne by PNIPAm microgels synthesized via FRP can affect their behavior in aqueous media~\cite{braibanti_impact_2016,truzzolillo_overcharging_2018} and in presence of oppositely charged polymers~\cite{truzzolillo_overcharging_2018}. Whether PNIPAm microgels synthesized via FRP behave like neutral colloids in water and how their VPT affects their electrostatics are still debated issues.}

In a recent work~\cite{truzzolillo_overcharging_2018} we have shown that PNIPAm microgels can electrostatically adsorb short polylysine polymers, that the amount of adsorbed polymer largely increases above the VPT and that such an adsorption triggers, {though only very close to the critical temperature}, both microgel reentrant condensation and overcharging, two phenomena characterizing charged colloids and occurring only in the presence of particles with sufficiently high charge densities.

\begin{figure}[htbp]
\centering
\includegraphics[width=10 cm]{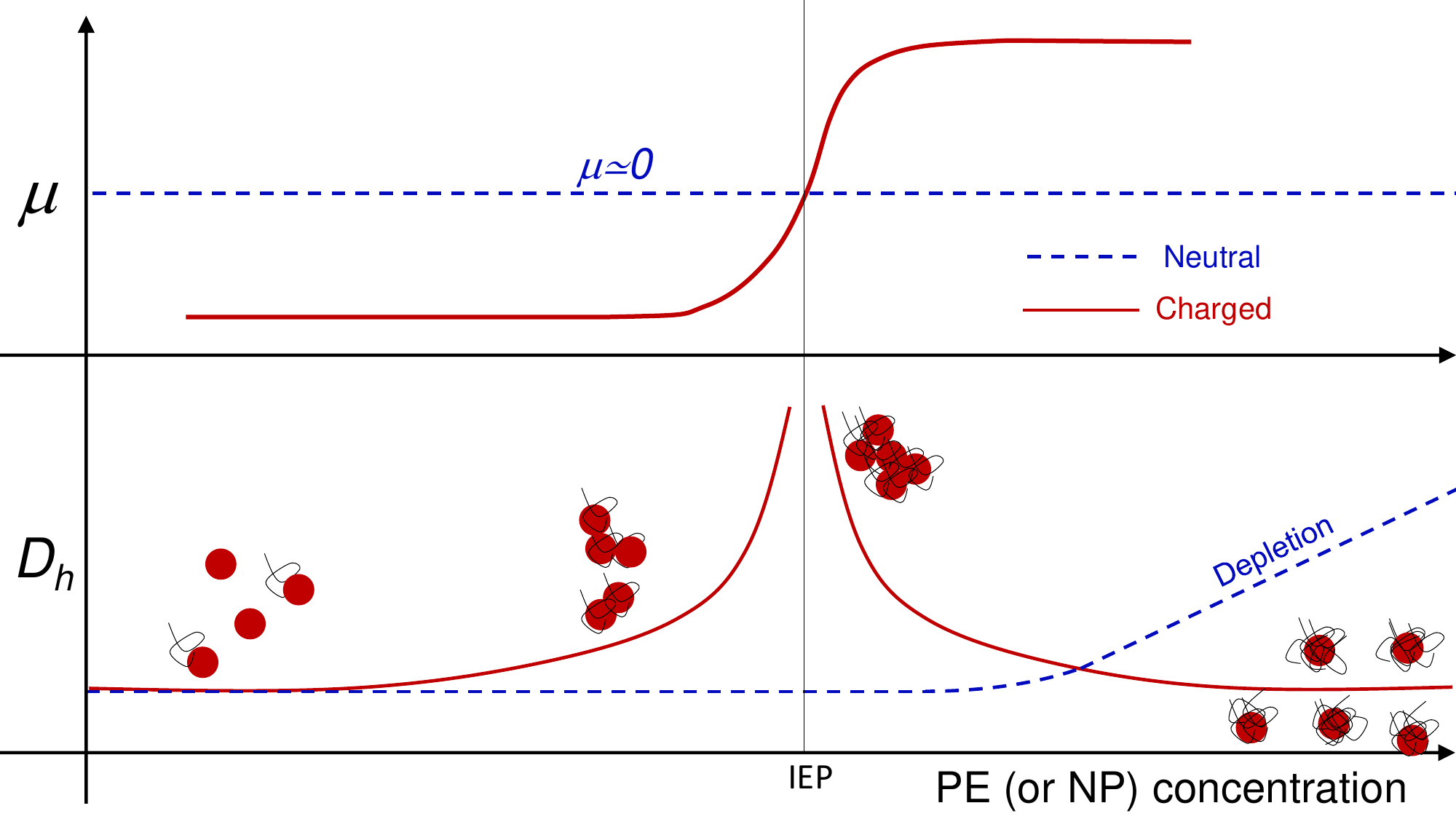}
\caption{Representative sketch of electrophoretic mobility ($\mu$) and hydrodynamic diameter ($D_h$) behavior when suspensions of neutral (dashed lines) and charged (continuous lines) colloids are progressively ``contaminated'' by small polymers or nanoparticles, whose charge is opposite to that of the colloids, when the latter also bear a non-zero net charge. Neutral colloidal particles are subject to a progressive clustering when depletion attractions dominate over thermal energy~\cite{lu_fluids_2006}, while their nearly zero mobility stays unaffected by depletant addition. Conversely, when charged polyelectrolytes or nanoparticles are added to a suspension of oppositely charged colloids both mobility reversal and reentrant condensation occur.}\label{sketch}
\end{figure}

Colloidal reentrant condensation and overcharging occur simultaneously (Figure~\ref{sketch}), when large multivalent ions, like polyelectrolytes (PE), adsorb onto oppositely charged colloids. By increasing the polyelectrolyte content, with the progressive reduction of the net charge of the primary polyelectrolyte-decorated particles, larger and larger clusters form. Close to the isoelectric point (IEP), where the charge of the adsorbed polyelectrolytes neutralizes the original charge of the particle surface, the aggregates reach their maximum size, while beyond this point any further increase of the polyelectrolyte-particle charge ratio causes the formation of aggregates whose size is progressively reduced. This re-entrant condensation behavior is accompanied by a significant overcharging. Overcharging, or charge inversion, occurs when more polyelectrolyte chains adsorb on a particle than are needed to neutralize its original charge so that, eventually, the sign of the net charge of the polymer-decorated particle is inverted. The stability of the finite-size long-lived clusters, that this aggregation process yields, results from a fine balance between long-range repulsive and short-range attractive interactions, both of electrostatic nature. For the latter, besides the ubiquitous dispersion forces, whose supply becomes relevant only at high ionic strength, the main contribution is due to the non-uniform correlated distribution of the charge on the surface of the polyelectrolyte-decorated particles (``charge-patch'' attraction)~\cite{bordi_polyelectrolyte-induced_2009}.

This complex phenomenology has been observed in a variety of polyelectrolyte-colloid systems dispersed in aqueous
solutions such as, for example, polyelectrolyte-micelle complexes~\cite{wang_polyelectrolytemicelle_2000}, latex particles~\cite{gillies_charging_2007,keren_microscopics_2002}, dendrimers~\cite{kabanov_interpolyelectrolyte_2000}, ferric oxide particles~\cite{milkova_complexation_2008}, phospholipid vesicles (liposomes)~\cite{bordi_direct_2006,bordi_polyelectrolyte-induced_2009,volodkin_complexation_2007,radler_structure_1998} and ``hybrid niosome'' vesicles~\cite{sennato_hybrid_2008}. Although polymer or nanoparticle adsorption can be further complicated by the presence of short-range interactions, specific to the different species, the similarities observed in such different systems strongly indicate that the overall phenomenology is mainly governed by non-specific electrostatic interactions, arising from double-layer overlap (repulsion) and surface charge non-uniformity (attraction). This makes electrostatic self-assembly of heterogeneously charged colloids profoundly different from depletion-driven aggregation observed in neutral colloid/polymer mixtures, where colloidal clustering is progressively favored by the addition small depletants (Figure~\ref{sketch}). {At present there is no clear experimental evidence for the existence of one type of colloid embedding both types of behavior, each of which potentially emerging at will by tuning a specific stimulus.}


In this well-established scenario, thermosensitive microgels, whose synthesis is initiated by ionic compounds, offer the unique possibility to tune colloidal charge density through thermal excitations. We have already demonstrated~\cite{truzzolillo_overcharging_2018} that, for short adsorbing $\varepsilon$-polylysine polymers with average molecular weight $M_w=$ 4 kD, the presence of stable clusters of PNIPAm microgels due to their overcharging, is limited by the weak adsorption energy ($\sim k_BT$) characterizing polyelectrolyte-microgel interaction. In this case, the high screening of electrostatic interactions due to the large number of free unadsorbed chains in the bulk and the importance of hydrophobic interactions between collapsed microgels, determined the occurrence of unstable clusters above the VPT, with a reentrant condensation detected only very close to the volume transition temperature ($T\sim$ 32--33 $^\circ$C), where swollen and collapsed microgels may coexist~\cite{polotsky_collapse--swelling_2013}.

In the present work we explore the onset of microgel aggregation and overcharging induced by the adsorption of large molecular weight ($M_w\geq$ 50 kD) polymers and one type of colloidal nanoparticles. {The properties of charged PNIPAm-based microgels covered by oppositely charged polyelectrolyte multilayers assembled through layer-by-layer procedures, have been thoroughly investigated in the literature~\cite{greinert_influence_2004,wong_layer-by-layer_2009}.
However, in these works the deposition of the polyelectrolyte layers is carried on at a fixed temperature, and as a consequence, at a fixed
charge density on the microgel surface.
Our work, highlighting the presence of a reentrant aggregation coupled to a large microgel overcharging when a temperature-stimulated gel shrinkage occurs and favors a subsequent electrostatic polyelectrolyte adsorption,
represents an important step forward towards a better understanding of these systems in view of technological applications.
We aim, in particular, at shedding light on the possibility to employ PNIPAm microgels to switch reversibly from neutral-like to charged-like colloids and on their capability to become heterogeneously charged (attractive) particles at high temperature with a well-defined zero-mobility point, where colloidal reentrant condensation takes place due to the adsorption of oppositely charged agents.}

By studying the complexation between microgels and oppositely charged lysine polymers via light scattering, electrophoresis, transmission electron microscopy (TEM) and atomic force microscopy (AFM), we show that cationic polylysine (PLL) chains adsorb onto PNIPAm microgels, causing their overcharging and reentrant condensation only above microgel VPT.
Such a phenomenology likewise appears in presence of anionic silica nanoparticles (NP) when cationic microgels are used.
Therefore we demonstrate, unambiguously for the first time, {that swollen PNIPAm microgel behavior conforms to that of neutral colloids}, while collapsed microgels show typical features of densely charged colloids. Comparing our present results with those presented in~\cite{truzzolillo_overcharging_2018}, we show that the molecular weight of polyelectrolytes, or rather, their overall charge and size, play an important role in determining the aforementioned phenomenology, large macromolecules being adsorbed more effectively than small ones, suggesting further the existence of a threshold molecular weight above which the PE adsorption does not depend on the chain length. 
By testing the persistence of such findings in a reverse charge system consisting of cationic microgels and anionic silica nanoparticles, we unveil the general character of this thermally driven electrostatic self-assembly.

The remainder of this paper is organized as follows. In Section~\ref{materials} we detail the materials and the experimental techniques employed to study PLL- and NP- microgels co-suspensions. Section~\ref{results} reports on the main results of our study and it is divided in three parts. The Section~\ref{barepnipam} describes the thermal behavior of bare PNIPAm microgels synthesized via surfactant-free FRP, and shows that the charge of the initiator determines the charge and the electrophoretic behavior of the microgels. Bare microgels are further visualized by AFM and their thermal behavior on solid support is displayed. The Section~\ref{microPLL} details the results obtained for co-suspensions of anionic PNIPAm microgels and cationic polylysine polymers. The presence of overcharging and reentrant condensation only above the microgel VPT, and for all the polyelectrolyte molecular weights will be pointed out. Polyelectrolyte adsorption on both swollen and collapsed microgels is further characterized by TEM and AFM imaging. The Section~\ref{microLudox} examines the phenomenology encountered in co-suspensions of cationic microgels and anionic silica nanoparticles, showing that overcharging and reentrant condensation characterize collapsed microgel behavior also in this case. The thermal reversibility of overcharged complex formation will be inspected in the Section~\ref{reversibility}, where we will discuss the presence of hysteresis when heating/cooling cycles are performed. Finally, in Section~\ref{conclusions} we make some concluding remarks and summarize the key results of this work.

\section{Materials and Methods}\label{materials}
\subsection{Materials}
PNIPAm microgels are synthesized by free-surfactant radical polymerization as detailed in~\cite{truzzolillo_overcharging_2018}.
We employed two ionic initiators, potassium peroxodisulfate (KPS) and 2,2'-Azobis(2-methylpropionamidine) dihydrochloride (AIBA), because of their proven efficacy in initiating FRP~\cite{fernandez-nieves_microgel_2011,chen_investigation_2017} and hence in determining the electrophoretic behavior of the final microgels.
The two initiators have generally two different initial reaction rates, as reported for other microgel systems~\cite{liu_initiator_2016}. This affects the size and the polydispersity index of the microgels. In particular, we expect that low reaction rates favor spatial heterogeneities and trigger the onset of broader size distributions, and viceversa. 
All syntheses have been performed at the same NIPAm concentration (0.01 wt/wt)~\cite{truzzolillo_overcharging_2018} in 150 mL of deionized water.
Chain crosslinking has been obtained through the addition of N,N-methylen-bis-acrylamide (BIS). Both NIPAm ad BIS have been primarily dissolved in 125~mL of deionized water, while initiators have been dissolved in 25~mL of water in a separate flask.
Solutions containing NIPAm and BIS have been bubbled with argon for 30 min and, after heating them up to 70 $^\circ$C, the initiator
solution has been added. After 6 h dispersions have been cooled down to room temperature and filtered through glass wool. NaN$_3$ (2 mmol) was
finally added to prevent bacterial growth.
We synthesized three types of microgels coded m1-KPS, m5-KPS and m5-AIBA, that are characterized by initiator-to-monomer and crosslinker-to-monomer molar ratios reported in Table~\ref{tablesynth}.

\begin{table}
\caption{Crosslinker/NIPAm and initiator/NIPAm molar ratios employed in FRP syntheses of the microgels.}\label{tablesynth}
\centering
\begin{tabular}{ccc}
\hline
\multicolumn{1}{l}{\small{\textbf{Sample code}}}	&\hspace{0.5 cm} \small{\textbf{Crosslinker/NIPAm molar ratio}}	& \hspace{0.5 cm}\small{\textbf{Initiator/NIPAm molar ratio}}\\
\hline
\multicolumn{1}{l}{m1-KPS}		& 0.013			& 0.016\\
\multicolumn{1}{l}{m5-KPS}		& 0.052			& 0.016\\
\multicolumn{1}{l}{m5-AIBA}		& 0.054			& 0.010\\
\hline
\end{tabular}
\end{table}

The m1-KPS and m5-KPS samples containing residual anionic persulfate (PS) groups have been further selected to study the complexation between microgels and cationic polyelectrolytes, while the m5-AIBA has been employed to investigate the complexation with anionic Ludox TM50 nanoparticles.

Charged groups are therefore embedded into the microgel structure and, since our syntheses are performed at $T=70$ $^{\circ}$C $>$ $T_{c}\sim$ 33 $^{\circ}$C, charges are preferentially located at the outer edge of the particles~\cite{zhou_correlation_2012}.

We have used 4 different hydrophilic polylysine polymers as adsorbing polymeric agents. Polylysine is an important class of polyamino acids with a broad spectrum of applications in biomedical research and development. It can be divided into two classes, $\alpha$-polylysine and $\varepsilon$-polylysine. The former is synthesized by artificial chemical synthesis and has limited applications due to its high toxicity, while the latter is produced by microbial synthesis and is widely used in various food, medicinal, and electronics products.

$\varepsilon$-Poly-L-lysine (coded as P$\varepsilon$4) was a kind gift from Chisso Corporation (Yokohama, Japan). This polymer, consisting of 25 to 35 L-lysine residues (Mw $\approx$ 4 kDa) is produced by a mutant of Streptomyces albulus NBRC14147 strain~\cite{Hiraki00,Hamano14}, and is used as a food preservative in several countries for its antimicrobial activity against a spectrum of microorganisms, including bacteria and fungi~\cite{Yoshida03}. $\varepsilon$-PLL is a hydrophilic cationic homo-poly-amino acid with an isoelectric point around pH = 9.0 and is described as having a peptide bond between carboxyl groups and $\varepsilon$-amino groups of L-lysine residues rather than the conventional peptide bonds linking $\alpha$-poly-L-lysine ($\alpha$-PLL)~\cite{hyldgaard_antimicrobial_2014} in which hydrophobic methylene side groups are fully exposed to water and may interact hydrophobically.

$\varepsilon$-PLL was in its basic form and was converted to Cl salt by titration with HCl followed by extensive dialysis to eliminate the H$^+$ excess.

\begin{figure}[htbp]
\centering
\includegraphics[width=3 cm]{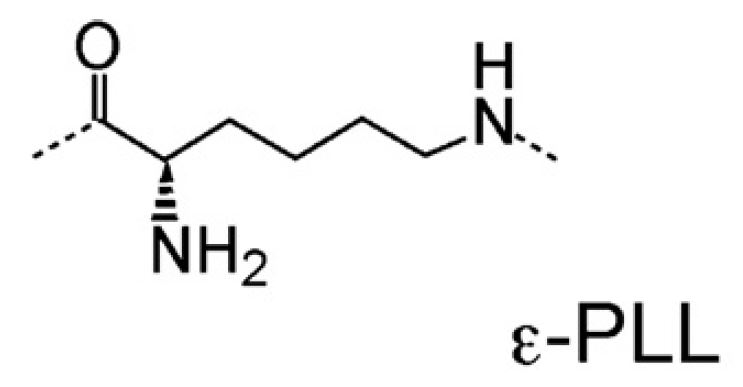}
\includegraphics[width=4 cm]{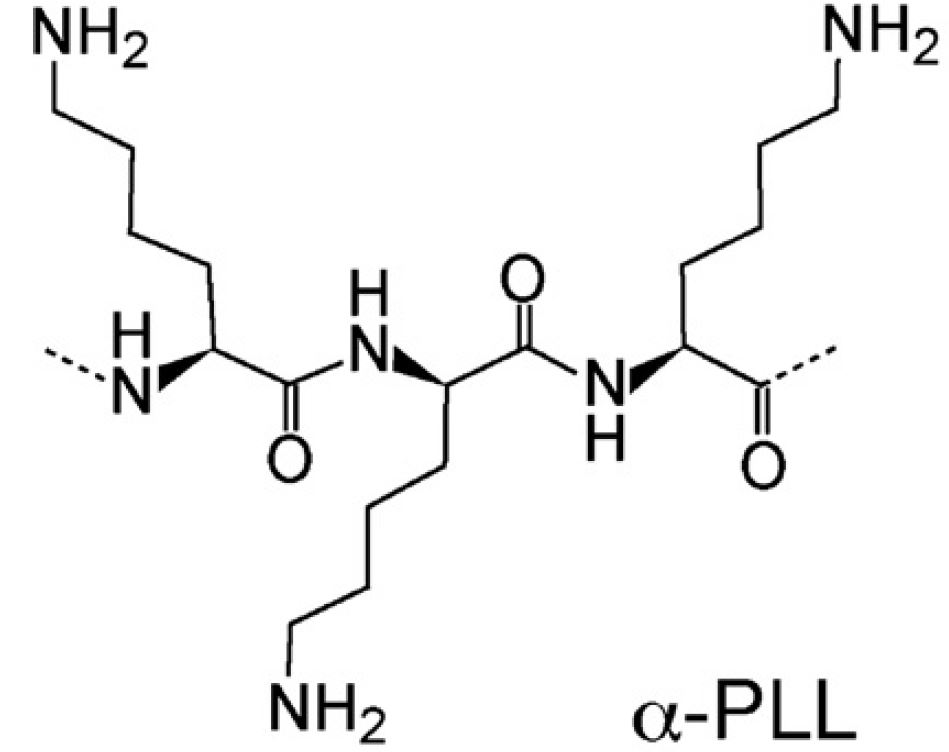}
\caption{Structure of polylysine polymers employed to prepare PLL-microgel mixtures. $\varepsilon$-PLL (\textbf{left}), a homopolymer consisting of the l-lysine residues linked by the peptide bonds between the $\varepsilon$-amino and the $\alpha$-carboxyl groups. $\alpha$-PLL (\textbf{right}), a homopolymer of l-lysine, whose structure is represented by three residues linked by peptide bonds between the $\alpha$-amino and $\alpha$-carboxyl groups. 
}\label{PLLstructure}
\end{figure}

Three $\alpha$-poly-L-lysine hydrobromide ($\alpha$-PLL) polymers with molecular weights $M_w=50$ kD (Code $P\alpha$50), $M_w=120$ kD (Code P$\alpha$120) and $M_w=250$ kD (Code $P\alpha$250) respectively were purchased from Polysciences, Inc. and used without any further purification. 
$\alpha$-PLL is the most well studied PLL polymer~\cite{leonard1963a,noguchi1966a,davidson1967a,ciferri1968a,myer1969a,greenfield1969a,liem1970a,pederson1971a,shepherd1976a,michael1989a,vorobjev1995a,dzwolak2004a}, and has been shown to adopt three different conformations ($\alpha$-helix, random coil and $\beta$-sheet structures) in aqueous solution depending on pH, temperature and additives (e.g., LiBr, MeOH and urea).
By contrast, conformational studies of ($\varepsilon$-PLL)~\cite{kushwaha1980a,maeda2003a,asano2007a,jia2010a}, being either of chemical~\cite{kushwaha1980a} or microbial origin~\cite{maeda2003a,asano2007a,jia2010a}, have reported its $\beta$-sheet structure in both solution and melt when the polymer is in the uncharged state, although a recent NMR study~\cite{jia2010a} showed that the microbial $\varepsilon$-PLL in the melt state (lyophilized at pH 5) is either in random coil or $\beta$-sheet structure depending on the molar mass. 
The two structures of the polymers used in this work are illustrated in Figure~\ref{PLLstructure}.

Finally, commercial Ludox-TM 50 anionic nanoparticles (NP) with hydrodynamic diameter $D_h=36$ nm~\cite{truzzolillo_bulk_2015-1} were purchased from Sigma-Aldrich and have been used without further purification as adsorbing agents co-suspended with m5-AIBA microgels.

\subsection{Preparation of microgel-polyion complexes}\label{preparation}
Each PLL-microgel and NP-microgel mixture was prepared according to the following standard protocol, which was well assessed in our past investigations on liposome-polyelectrolyte complexes (see for example~\cite{sennato_salt-induced_2016}).
A volume of 0.5~mL of a PLL (or NP) solution at the required concentration was added to an equal volume of the microgel suspension in a single mixing step and gently agitated by hand. Before mixing, both microgel and polyelectrolyte (or nanoparticle) suspensions were kept at room temperature to avoid interference of thermal gradients during the subsequent measurement. After mixing the two components, the sample was immediately placed in the thermostatted cell holder of the instrument for the measurement of the electrophoretic mobility and the size of the resulting complexes.

\subsection{Viscosimetry}\label{visco}
Viscosity measurements were performed using an Anton Paar Lovis 2000 ME micro-visco\-si\-meter to obtain the constant of proportionality between PNIPAm mass fraction, $c_m$, and microgel volume fraction, $\varphi$, at $T=20$ $^{\circ}$C.
We remind that PNIPAm microgels are soft colloids whose dynamics is dictated by their effective volume fraction $\varphi$~\cite{philippe_glass_2018} and the interparticle potential. It is then crucial to measure $\varphi$ to avoid particle crowding or gelation at high temperature~\cite{romeo_temperature-controlled_2010}, since they may affect particle mobility and the correct measurements of particle size via dynamic light scattering.
In the range $6.25\times 10^{-5} < c_m < 7.48 \times 10^{-4}$ the viscosity $\eta$ of the suspensions increases linearly with $c_m$. Since microgels are highly swollen, their mass density is essentially the same as that of the solvent. Consequently, weight fraction $c_m$ and volume fraction $\varphi$ are proportional, i.e., $\varphi = kc$. We determined the constant $k$ using the $c$-dependence of the zero-shear viscosity in the dilute regime~\cite{truzzolillo_bulk_2015-1}. Briefly, we determined the constant $k$ by matching the concentration dependence of the zero shear viscosity to the one predicted in the dilute regime by Einstein's formula:
\begin{equation}\label{eta}
\frac{\eta}{\eta_0}=1+2.5\varphi=1+2.5kc_m
\end{equation}
where $\eta_0$ is the viscosity of the solvent. By fitting $\eta/\eta_0$ to a straight line, we obtained $k = 23.9 \pm 1.3$, $k = 12.1 \pm 0.4$ and $k = 14 \pm 1$ for m1-KPS, m5-KPS and m5-AIBA respectively.
This allows defining the microgel volume fraction as $\varphi (T)=kc$ $D_h^3(T)$/$D_h^3(20~^{\circ}$C$)$, where $D_h(T)$ is the hydrodynamic diameter of the microgels measured by dynamic light scattering. All the suspensions and mixtures containing anionic m1-KPS and m5-KPS particles have been prepared at fixed microgel number density correspondent to $\varphi$(20 $^{\circ}$C) $= 0.024$, while m5-AIBA cationic microgels, employed to prepare mixtures with silica NP, have been used at a fixed concentration $c_m=0.001$ wt/wt correspondent to $\varphi$(20 $^{\circ}$C) $= 0.012$. For such fractions the self diffusion coefficient coincides with the one measured in the very dilute limit~\cite{philippe_glass_2018} and hence the hydrodynamic size could be properly measured. 

Additional steady rate rheology experiments, using a stress-controlled AR 2000 rheometer (TA Instruments), equipped with a steel cone-and-plate geometry (cone diameter = 50 mm, cone angle = 0.0198 rad) have been performed to measure the zero shear viscosity of aqueous samples of $P\alpha$250 at 1 mg/mL and 0.5 mg/mL in the temperature range $20$ $^{\circ}$C $\leq T\leq 45$ $^{\circ}$C and shear rate ranging from 1000 s$^{-1}$ down to 1 s$^{-1}$. The viscosity ranged from 2.30 mPas (at 20 $^{\circ}$C) down to 1.90 mPas (at 45 $^{\circ}$C) for the sample at 1~mg/mL, while at 0.5 mg/mL the viscosity ranged from 1.73 mPas down to 1.69 mPas in the same temperature range. All samples showed Newtonian behavior, since viscosities did not change appreciably by varying the imposed shear rate.
All the other samples were characterized by viscosities deviating less than 3\% from the one of deionized water and hence they have been considered isoviscous with respect to the pure solvent. Such viscosity values have been successively used to extract the hydrodynamic size of PLL- and NP-microgels complexes or clusters via DLS.

\subsection{Light scattering and Electrophoretic measurements}\label{dls}
The hydrodynamic size and the size distribution of microgels and microgel-based complexes were characterized by means of dynamic light scattering (DLS) measurements, employing a MALVERN Nano Zetasizer apparatus equipped with a 5 mW HeNe laser (Malvern Instruments LTD, UK). The scattered light is collected at an angle of 173$^{\circ}$. The main advantage of this detection geometry, when compared to the more conventional detection at 90$^{\circ}$, is its inherent larger insensitiveness to multiple scattering effects~\cite{dhadwal_fiberoptic_1991}. Moreover, as large particles scatter mainly in the forward direction, the effects on the size distribution of dust or, as is our case, of large irregular aggregates (lumps or clots),  are greatly reduced. To obtain the mean size $D_h$ and the polydispersity index (PDI), we performed a cumulant analysis of the measured intensity autocorrelation functions $g_2(\tau)-1$ up to the second central moment of the distribution of decay rates~\cite{berne_dynamic_1990}. To further probe the presence of free (non-adsorbed) polyelectrolytes and nanoparticles in PE- and NP- microgel suspensions respectively, autocorrelation functions have been also analyzed by means of the CONTIN algorithm~\cite{provencher_constrained_1982} trough which we extracted the intensity-weighted size distributions. 
For the sake of clarity, we recall here that the intensity autocorrelation functions at one fixed scattering vector are calculated as follows
\begin{equation}\label{intensitycorr}
    g_2(\tau)-1=\frac{\langle I(\tau+t)I(t)\rangle}{\langle I(t)\rangle^2},
\end{equation}
where $I$ is the intensity of the light scattered by the sample and $\tau$ is the delay time between two distinct intensity detection. $g_2(\tau)-1$ are exponentially decaying functions for samples containing diffusive Brownian particles and the decay times are used to determine the average diffusion coefficients $\tilde{D}$ of the particles, which in turn can be converted in hydrodynamic diameters, $D_h$, using the Stokes-Einstein equation $D_h = k_BT/3\pi\eta \tilde{D}$, where $k_B$ is the Boltzmann constant, $T$ the absolute temperature and $\eta$ the solvent viscosity~\cite{berne_dynamic_1990}.\\ The values of $D_h$ shown in this work correspond to an average over 3 independent measurements.

The electrophoretic mobility has been measured by means of the same NanoZetaSizer apparatus. This instrument is integrated with a laser Doppler electrophoresis technique, and the particle size and electrophoretic mobility can be measured almost simultaneously and in the same cuvette. In this way, possible experimental uncertainties due to different sample preparations, thermal gradients and convection are significantly reduced. The average electrophoretic mobility is determined using the Phase Analysis Light Scattering (PALS) technique~\cite{tscharnuter_mobility_2001}, a method which is useful especially at high ionic strengths, where mobilities are low. In these cases the PALS configuration has been shown to be able to measure mobilities two orders of magnitudes lower than traditional light scattering methods based on the shifted frequency spectrum (spectral analysis).

We adopt a thermal protocol consisting of an ascending ramp from 20 $^{\circ}$C to 45 $^{\circ}$C with temperature step of 1 $^{\circ}$C. At each step, samples have been left to thermalize 300 s, then measurements of electrophoretic mobility and size have been performed.
\subsection{Microscopy}\label{Mic}
Morphology of bare m1-KPS microgels and P$\alpha$250/m1-KPS complexes were studied by Transmission Electron Microscopy (TEM) and Atomic Force Microscoy (AFM). Samples for TEM measurements have been prepared by depositing 20 $\mu$L of suspensions  on a 300-mesh copper grid for electron microscopy covered by a thin amorphous carbon film.   After 5 min the excess liquid was removed by touching the grid to filter paper and 10~$\mu$L of 2 $\%$ aqueous phosphotungstic acid (PTA)  (pH-adjusted to 7.3 using 1 N NaOH) has been added to stain the sample.
To investigate morphological variations induced by temperature, samples have been deposited both at room temperature and 40 $^\circ$C.
In the latter case, in order to avoid thermal gradients, TEM grids, pipette tips and PTA solution have been heated at 40 $^\circ$C and both deposition and drying have been carried out in a thermostatted oven. Bare m1-KPS suspensions have been thermalized  for 30 min at 40 $^\circ$C and then deposited on the pre-heated grid.  P$\alpha$250/m1-KPS samples have been mixed at room temperature and then heated up to 40 $^{\circ}$C  according to the thermal protocol described above.
Measurements were performed by a FEI TECNAI 12 G2 Twin (Thermo Fisher Scientific-FEI Company, Hillsboro, OR, USA), operating at 120 kV and equipped with an electron energy loss filter (Biofilter, Gatan Inc, Pleasanton, CA, USA) and a slow-scan charge-coupled device camera (794 IF, Gatan Inc, Pleasanton, CA, USA).

For Atomic Force Microscopy (AFM) measurements, P$\alpha$250/m1-KPS samples have been deposited on freshly cleaved mica support after dilution (50$\times$). Samples have been deposited both at room temperature and 40 $^\circ$C by using the same sample preparation protocol described for TEM. Measurements were performed by a Dimension Icon (Bruker AXS). AFM images were acquired in air at room temperature and under ambient conditions by employing Tapping Mode and RTESP-300 (Bruker) probes (nominal radius of curvature 10 nm). Mica discs were fixed on the sample support of the instrument by using double-sided tape 1h before the measurement to allow sample stabilization and avoid mechanical noise. AFM data processing have been performed by Gwiddion software. Background subtraction through line-by-line or three-point data leveling of height sensor channel data has been considered.
\section{Results}\label{results}
\subsection{Characterization of bare microgels}\label{barepnipam}
\begin{figure}[htbp]
\centering
\includegraphics[width=13 cm]{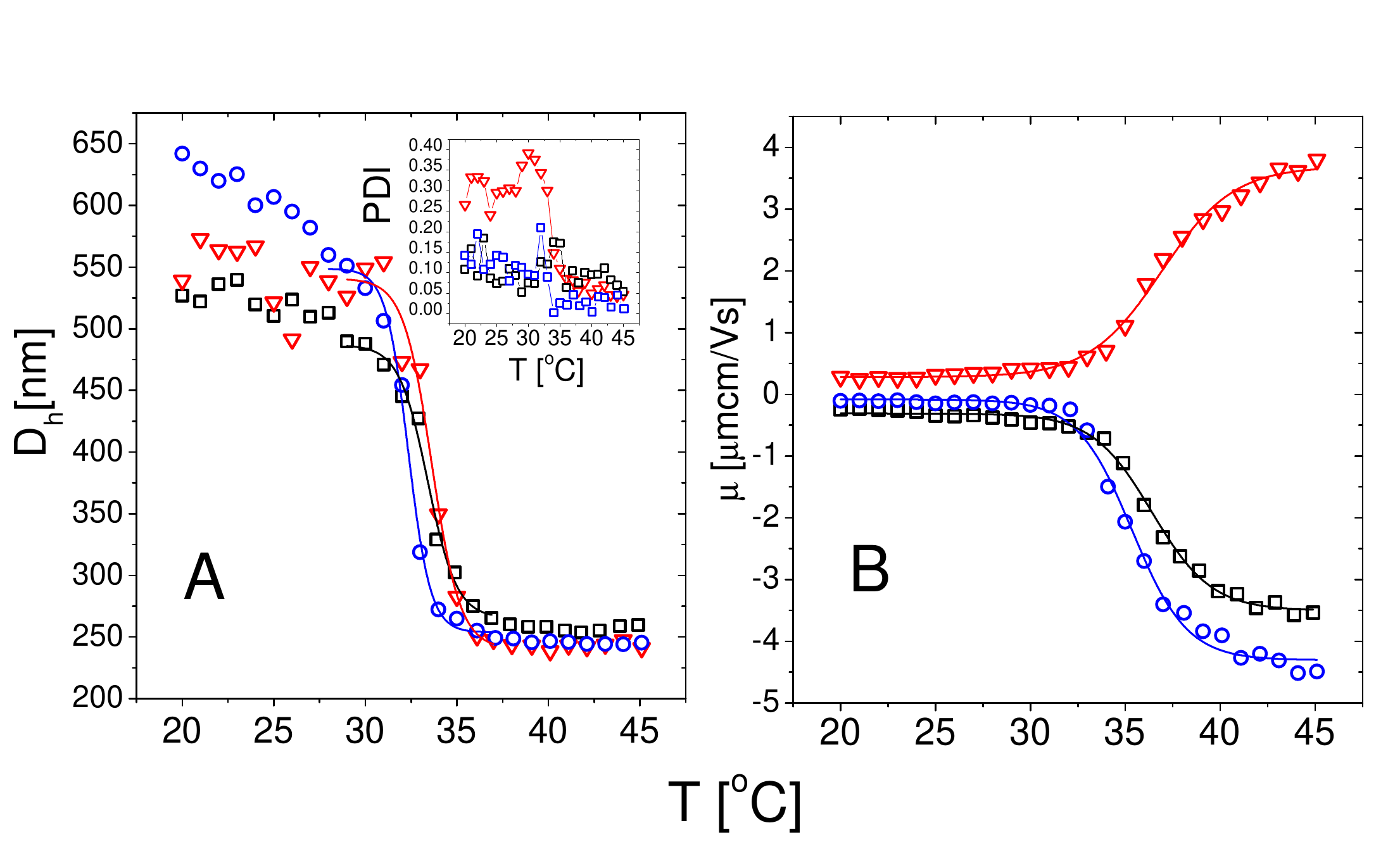}
\caption{Hydrodynamic diameter (Panel \textbf{A}) and electrophoretic mobility (Panel \textbf{B}) of m1-KPS (circles), m5-KPS (squares) and m5-AIBA (triangles) microgels in water as a function of temperature. Solid lines are fits obtained using Boltzmann functions as detailed in the text. The inset (Panel \textbf{A}) shows the polydispersity index (PDI) obtained via a cumulant analysis of the intensity autocorrelation functions $g_2(\tau)-1$ for the three microgel types.}\label{ancat} 
\end{figure}

Microgel size, as measured by $D_h$, abruptly decreases as the temperature is raised above PNIPAm LCST (Figure~\ref{ancat}). For all microgels the diameter displays a characteristic sigmoidal behavior for increasing temperatures, that is the fingerprint of the microgel VPT and has been widely discussed in literature~\cite{pelton_particle_1989,maki_temperature_2018,braibanti_impact_2016,truzzolillo_overcharging_2018}. The volume phase transition temperatures have been extracted by fitting the data in the temperature range $29$ $^\circ$C $\leq T \leq$ $37$~$^\circ$C via a Boltzmann-type nonlinear regression $D_h(T)=D_{\infty}+(D_0-D_{\infty})/\left[1+e^{(T-T_{c})/dT}\right]$ and corresponds to the flex point of the curve. The parameter $dT$ further characterizes the sharpness of the transition. Indeed, since the normalized slope at the critical temperature is $s_c=|D_h'(T_c)|/|D_{\infty}-D_0|=1/(4dT)$, we can compute the latter to characterize the temperature dependence of the particle size across the transition. The best-fit parameters are reported in Table~\ref{tableboltz}.

Although the microgels show a very similar swelling-deswelling behavior, m5-AIBA microgels are characterized by a larger polydispersity index below $T_c$ with respect to m1-KPS and m5-KPS particles (Figure~\ref{ancat}A---Inset). This might be due a lower reaction rate of AIBA initiator as reported for other microgel systems~\cite{liu_initiator_2016}. The drop of the PDI above $T_c$ for m5-AIBA microgels points to particles characterized by a rather heterogenous structure made of a compact core and more polydisperse loose peripheral chains~\cite{braibanti_impact_2016}. By collapsing, microgels homogenize their structure and, by extension, their size distribution.

Figure~\ref{ancat}B shows the average electrophoretic mobility $\mu(T)$ for m1-KPS, m5-KPS and m5-AIBA suspensions as a function of the temperature at $\varphi(20~^{\circ}$C$)=0.024$, which is also the microgel volume fraction used for the particle size and the electrophoretic characterization of PLL-microgel mixtures. $\mu(T)$ is affected by the VPT and increases in magnitude as the temperature is increased above $T_c$. However, for $T=T_c$ the mobility magnitude is only $\approx$2 times larger than its value at 20 $^{\circ}$C for all the microgel types and the mobility increase is somehow shifted with respect to the drop in size.
Similarly to what has been done for the microgel size, we quantify the mobility drop by fitting the data over the whole temperature range $20$ $^\circ$C$\leq T \leq$$45$~$^\circ$C via a Boltzmann-like function $\mu(T)=\mu_{\infty}+(\mu_0-\mu_{\infty})/\left[1+e^{(T-T_{c\mu})/dT_{\mu}}\right]$. This allowed us defining an 'electrokinetic transition temperature' $T_{c\mu}$ and a normalized slope at the transition $s_{c\mu}=|\mu'(T_c)|/|\mu_{\infty}-\mu_0|=1/(4dT_{\mu})$ (Table~\ref{tableboltz}).


It is worth noting that the difference between the pairs ($T_c$, $s_c$) and ($T_{c\mu}$, $s_{c\mu}$) is significant, suggesting that the electrokinetic transition is both delayed and smoother in temperature with respect to the volume one for all the adopted FRP synthesis protocols.

Such a significant difference between these two transitions has been already discussed by Pelton et al.~\cite{pelton_particle_1989} and by Daly et al.~\cite{daly_temperaturedependent_2000} and has been attributed to a multi-step process, where the almost-uncharged core collapses first, with a significant reduction of particle size, and the shell, where the charges are mostly confined, collapsing only at higher temperature. Moreover, the opposite sign of the electrophoretic mobility of microgels whose polymerization has been initiated by KPS and AIBA demonstrates quite unambiguously that the charge of a PNIPAm microgel is dictated solely by the ionized groups of the initiator molecules. Therefore the latter can be used to tune microgel charge density and sign. 

The amount of crosslinker also has an effect, albeit weak, on both the volume and electrokinetic transition: m1-KPS sample is characterized by lower transition temperatures ($T_c$, $T_{c\mu}$) and higher slopes ($s_{c\mu}$, $s_{c\mu}$) with respect to the other two more crosslinked microgels, suggesting that lowering crosslinker density hinders less polymer collapse, favoring its occurrence at lower temperatures. This is also corroborated by the continuous increase of the microgel size upon decreasing temperature below $T_c$ for the m1-KPS particles, that signals the presence of loose peripheral PNIPAm chains in this type of microgels. In line with this finding, m1-KPS microgels show also the widest change of electrophoretic mobility (Figure~\ref{ancat}B) versus temperature, as evidenced by the largest limiting mobility difference $|\mu_{\infty}-\mu_{0}|$ obtained for the synthesized microgels (Table~\ref{tableboltz}), pointing to a larger variation of the charge distribution within the networks as temperature varies.
\begin{table}
\caption{Volume phase transition temperatures ($T_c$), electrokinetic transition temperatures $T_{c\mu}$, relative normalized slopes ($s_c$ and $s_{c\mu}$) and limiting mobility difference $|\mu_{\infty}$-$\mu_{0}|$ obtained via Boltzmann fits as detailed in the text.}\label{tableboltz}
\centering
\begin{tabular}{cccccc}
\hline
\textbf{Sample code} &\hspace{0.5 cm} \textbf{$T_c$} [$^\circ$C] &\hspace{0.5 cm} \textbf{$s_c$} [$^\circ$C$^{-1}$] &\hspace{0.5 cm} \textbf{$T_{c\mu}$} [$^\circ$C] &\hspace{0.5 cm} \textbf{$s_{c\mu}$}[$^\circ$C$^{-1}$] &\hspace{0.5 cm} $|\mu_{\infty}$-$\mu_{0}|$ [$\mu$m cm/Vs]\\
\hline
m1-KPS		&\hspace{0.5 cm} $32.3\pm0.1$ &\hspace{0.5 cm}	$0.42\pm0.05$	&\hspace{0.5 cm} $35.3\pm0.1$ &\hspace{0.5 cm} $0.17 \pm 0.01$ &\hspace{0.5 cm} $4.3 \pm 0.1$\\
m5-KPS		&\hspace{0.5 cm} $33.4\pm0.1$ 	&\hspace{0.5 cm} $0.32 \pm 0.06$	&\hspace{0.5 cm} $36.4 \pm 0.1$ &\hspace{0.5 cm} $0.16 \pm 0.01$ &\hspace{0.5 cm} $3.19 \pm 0.06$\\
m5-AIBA		&\hspace{0.5 cm} $33.60\pm0.15$ &\hspace{0.5 cm}	$0.32 \pm 0.09$   &\hspace{0.5 cm} $36.9 \pm 0.1$ &\hspace{0.5 cm} $0.13 \pm 0.01$ &\hspace{0.5 cm} $3.41 \pm 0.09$\\
\hline
\end{tabular}
\end{table}

\begin{figure}[htbp]
\centering
\includegraphics[width=13 cm]{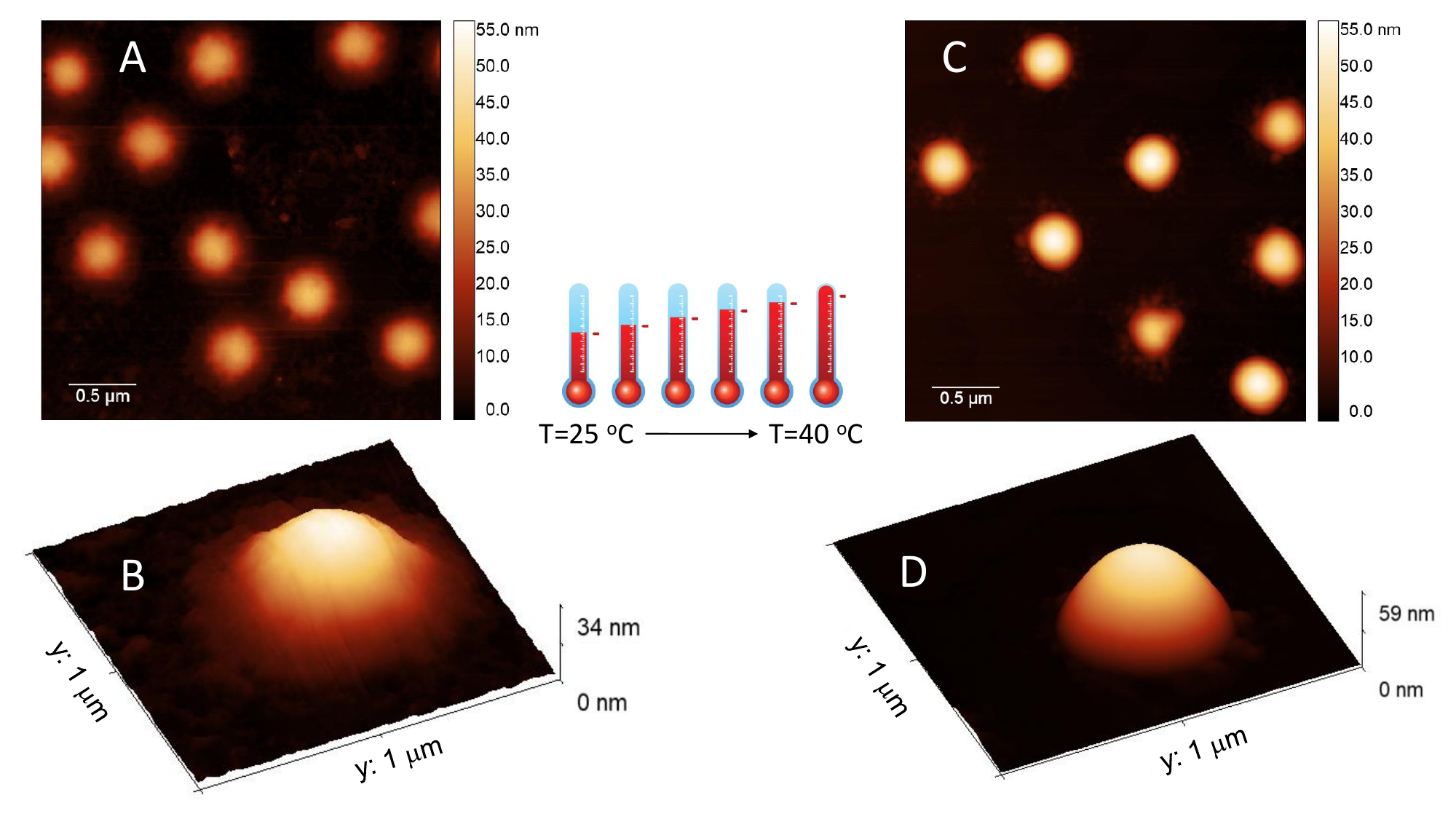}
\caption{AFM height images and single-particle 3d-projection of bare m1-KPS PNIPAm microgels deposited at 25~$^\circ$C (Panels \textbf{A} and \textbf{B}) and 40 $^\circ$C (Panels \textbf{C} and \textbf{D}). The different appearance of PNIPAm particles is determined by the variation of  the hydrophilic character of the polymer network, which influences the interaction with the hydrophilic mica substrate. At T = 40 $^{\circ}$C microgel particles become less affine to the substrate, with respect to room temperature, so dried particles result more compact and spherical.}\label{AFMbare}
\end{figure}

Finally Figure~\ref{AFMbare} shows AFM images of m1-KPS PNIPAm microgels deposited at 25~$^\circ$C (Panels A and B) and 40 $^\circ$C (Panels C and D) and measured in air. The morphological differences beetween the two swollen and collapsed states  of microgels are evident, as expected on the basis of previous investigations~\cite{tagit_probing_2008}. At 25 $^\circ$C microgels sit on the support in a pancake-like shape with the poorly-crosslinked periphery laying expanded due to the affinity with the hydrophilic surface of mica. Similarly to what observed by Tagit and coworkers~\cite{tagit_probing_2008}, dried particles exhibit a flattened shape with average height values almost ten times smaller than their (apparent) width when deposited below the VPT.
By contrast, at 40 $^\circ$C microgels  show a more compact and spherical morphology, with a reduction of the apparent particle surface area and an increase of height due to the unfavorable interaction with the hydrophilic support, which is evident also in dried conditions. These images represent a frame of reference for the more complex behavior of PLL-microgel complexes that will be discussed hereafter.

\subsection{Polyelectrolyte-Microgel Suspensions}\label{microPLL}
\begin{figure}[htbp]
\centering
\includegraphics[width=13 cm]{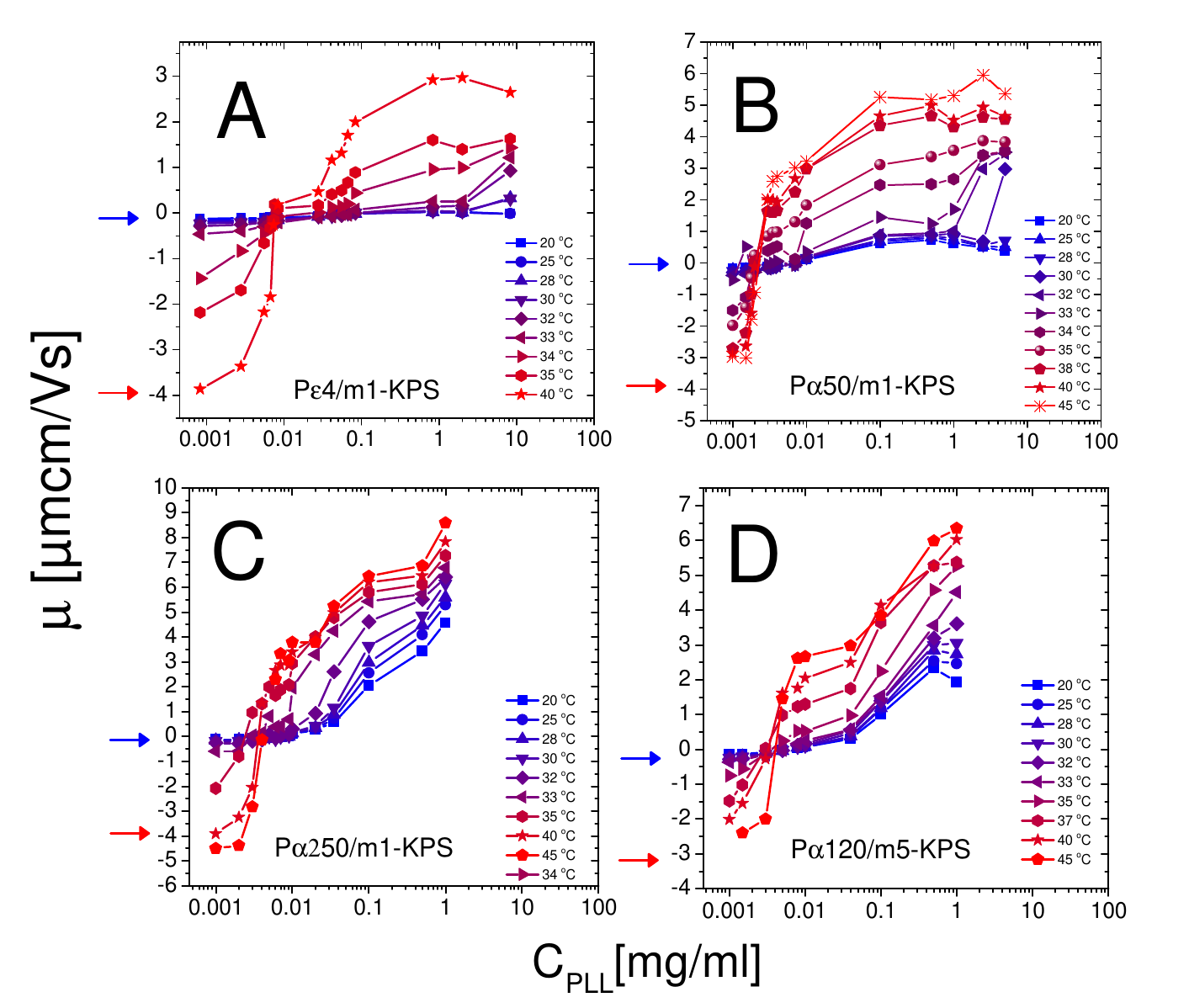}
\caption{Electrophoretic mobility in function of $C_{PLL}$ for P$\varepsilon$4/m1-KPS (Panel \textbf{A}), P$\alpha$50/m1-KPS (Panel \textbf{B}), P$\alpha$250/m1-KPS (Panel \textbf{C}) and P$\alpha$120/m5-KPS (Panel \textbf{D}) mixtures, and selected temperatures as indicated in the panels. Mobilities of bare microgels at 20 $^\circ$C (blue arrows) and 40 $^\circ$C (red arrows) are also shown for reference.} \label{mobilityall}
\end{figure}

In this section we report a detailed inspection on the polyelectrolyte-microgel complexation and explore the possibility that the augmented charge density of the microgels above $T_{c\mu}$ gives rise to the phenomenology sketched in Figure~\ref{sketch} (red lines), characterizing most of the colloids exposing to the solvent densely distributed ionizable groups.
Figure~\ref{mobilityall} shows the electrophoretic mobility of anionic microgels (m1-KPS and m5-KPS) in presence of PLL polymers of different molecular weight, structure, and varying temperature and PLL concentration. We have both tested the role of PLL molecular weight by fixing the microgel type (m1-KPS) and changing the PLL length, and, albeit only marginally, the role of the crosslinker density.

Two features of the electrophoretic behavior of these mixtures must be primarily emphasized.

\begin{figure}[htbp]
\centering
\includegraphics[width=12 cm]{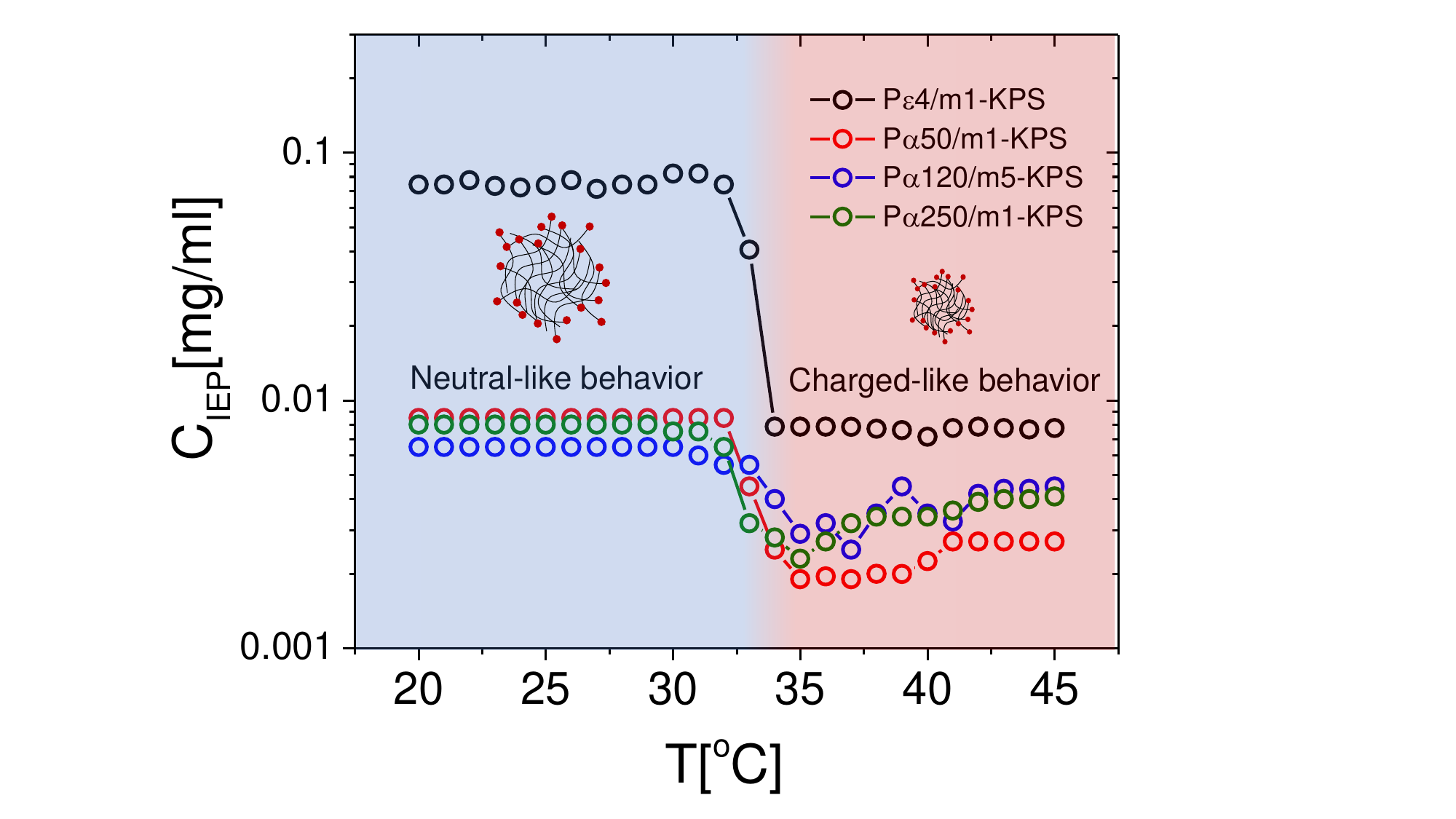}
\caption{$C_{IEP}$ in function of temperature for all the investigated PLL-microgel mixtures. The sudden drop of $C_{IEP}$ in proximity of $T_c$ marks the electrokinetic transition between the neutral-like and the charged-like state of PNIPAm microgels.}\label{IEP}
\end{figure}

\begin{itemize}
\item[(i)] Though we observe systematically the crossover from negative to positive average mobilities for all temperatures, the modalities with which such a variation occurs change qualitatively when passing from $T \lesssim T_c$ to $T \gtrsim T_c$. Below $T_c$ mobility smoothly increases, without showing a sudden jump from negative to positive values and suggests that PLL chains adsorb onto PNIPAm microgels even when the latter are swollen. Above $T_c$ the mobility passes from highly negative to highly positive values in a very narrow range of PLL concentrations. This is what is expected when ion-ion correlation increases in the Debye layer in proximity of an adsorption surface.
In fact, based on a modified Poisson equation, proposed by Bazant and coworkers~\cite{bazant_double_2011} to capture the effect of ion-ion correlation in equilibrium Debye layers, Stout and Khair~\cite{stout_continuum_2014} showed recently that the extent of electrophoretic mobility reversal and its sharpness in colloid-multivalent ion systems are both enhanced by an increase of the ion-ion correlation length. In the case of PNIPAm microgels with tunable charge density we attribute this increased ion-ion correlation to the enhanced adsorption energy on the microgel periphery.  Moreover, the authors~\cite{stout_continuum_2014} find that the point where $\mu$ vanishes does not correspond exactly to the point of zero charge, and that the mobility reversal occurs at progressively lower ion concentration for increasing correlation length among the ions. As a matter of fact, when a colloid is immersed in a large multivalent ion solution a large contribution to the net force exerted on the colloid is given by the electro-osmotic hydrodynamic force, constantly directed in opposite direction to the applied electric field.

Both electric and electro-osmotic force decrease in magnitude as the colloid is progressively neutralized by spatially correlated polyions, but the rate of this decrease differs for the two forces as the correlation length characterizing adsorbed polyion is large enough. For such a reason a still negatively charged decorated colloid can move in the electric field direction and hence with a reversed mobility and the original sign of the charge.
{Though the different porosity and electrolyte diffusion time scales characterizing swollen and collapsed microgels should be considered to exhaustively capture the physics ruling the microgel electrophoretic behavior~\cite{liang_fractal_2019}, this overall scenario conforms to our polyelectrolyte-microgel complexes, where polyion correlation is intensified by an increase of the adsorption energy due to microgel collapse.} This said, for our systems the isoelectric point, classically identified as the point where $\mu=0$, does not lie far from the true isoelectric condition, where the net charge of the complex is zero. As we will show later in this section, large aggregates form and flocculation occurs in proximity of the mobility reversal, where thus the net charge of decorated microgels must be low and the interaction among them dominated by charge patch attractions.

\item[(ii)] The PLL concentration $C_{IEP}$ at which the average mobility reversal occurs sharply drops at $T \backsimeq T_c$ (data reported in Figure~\ref{IEP} and further discussed later in this section).

\end{itemize}
Secondarily, we also detect no evident effect of microgel softness in the investigated range of crosslinker-to-monomer molar ratio: P$\alpha$120/m5-KPS mixtures do not show different electrokinetic features with respect to the others, where the m1-KPS particles have been employed. This has to be expected since m1-KPS and m5-KPS microgels show very similar values of ($T_c$, $s_c$) and ($T_{c\mu}$,$s_{c\mu}$) (Table~\ref{tableboltz}) and show comparable size reduction for $T>T_c$. We infer that a much larger crosslinker density must be considered to investigate the role played by microgel softness in polyelectrolyte adsorption. In this regard we expect a suppression of the reentrant condensation of microgels when their softness is sufficiently lowered, since a big variation of their charge density is hampered by the reduced extent of the VPT. On that note, we wish to reiterate here that the study of P$\alpha$120/m5-KPS suspension has served as a probe for the universal and  non-specific character of the phenomenology encountered in mixtures of microgels and polyelectrolytes and it was not thought as a way to investigate exhaustively the effect of softness.

So, all in all, our findings are in line with m1-KPS and m5-KPS particles undergoing an electrokinetic transition from neutral-like to charged-like colloids.

\begin{figure}[htbp]
\centering
\includegraphics[width=13 cm]{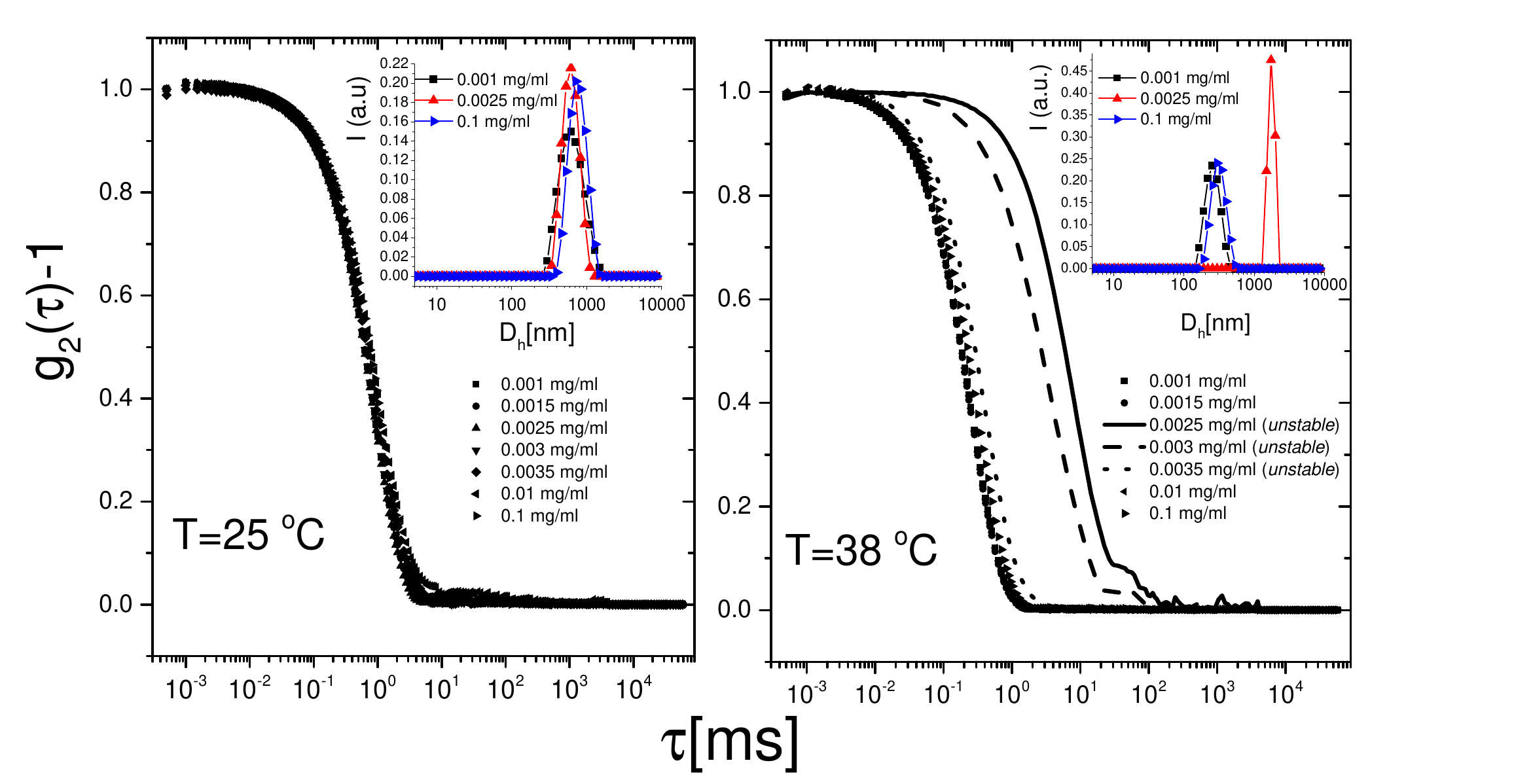}
\caption{Intensity autocorrelation functions $g_2(\tau)$-1 for P$\alpha$50/m1-KPS mixtures at selected $C_{PLL}$ (see panels) and two temperatures, 25 $^{\circ}$C (Left panel) and 38~$^{\circ}$C (Right panel), representing the neutral-like and the charged-like behavior of PNIPAm microgels. A reentrant shift to much longer relaxation times of the correlograms appears only at high temperatures ($T>T_c$), while it is absent for $T<T_c$. The inset panels show the intensity-weighted size distributions obtained by CONTIN analysis of the correlograms for three selected P$\alpha$50 concentrations, correspondent to poorly P$\alpha$50-decorated microgels, nearly neutralized microgels and overcharged microgels.}\label{g2-panel}
\end{figure}

The self-assembly behavior of the mixtures corroborates this scenario. Figure~\ref{g2-panel} shows the intensity autocorrelation functions
$g_2(\tau)$-1 measured via DLS as described in Section~\ref{dls} for selected $P\alpha$50/m1-KPS mixtures, at different $P\alpha$50 concentrations and two temperatures: 25 $^{\circ}$C (Panel A) and 38 $^{\circ}$C (Panel B), i.e., below and above $T_{c}$ respectively. Below $T_{c}$ the correlograms do not show any significant variation for increasing PLL concentrations, but for a very small shift towards larger delay times. This presumably results from a partial adsorption of the PLL chains, causing an overall increase of the size of PLL-microgel complexes without affecting the stability of the suspensions. By contrast, above $T_{c}$, correlation functions do show a striking dependence on PLL concentration. A very large shift to longer lag times is detected only in a narrow range of PLL concentrations, for which macroscopic flocs are detected and correlograms are unsteady: microgels undergo reentrant clustering at high temperature.
To inspect whether the collected light was partially scattered by free (non-adsorbed) PLL chains in the bulk, we preliminarily performed a CONTIN analysis of the correlograms. The inset panels of Figure~\ref{g2-panel} show the intensity-weighted size distribution at three concentrations of $P\alpha$50, corresponding to: (i) very low polyion content (0.001 mg/mL), where mobilities are close to the one of bare microgels; (ii) a moderate concentration (0.0025 mg/mL), where the average mobility value is close to zero while it sharply changes upon increasing $C_{PLL}$ for $T>T_c$; (iii) a high PLL concentration (0.1~mg/mL) where microgels are overcharged at any $T$ (Figure~\ref{mobilityall}).
We find unimodal intensity-weighted size distributions in all cases, suggesting that light scattered by all samples is dominated by PLL-microgel complexes (or clusters), most of the chains are adsorbed on the microgels, and hence the residual free chains do not scatter enough to be detected. We recall here that NiPam and PLL polymers are characterized by two different refractive indices: the values $n_{pnipam}=$ 1.52~\cite{li_submicrometric_2015} and $n_{PLL}=$ 1.37~\cite{wang_synthesis_2003} have been reported for PNIPAm chains in water and grafted random coil PLL polymers respectively. Therefore the detection few residual PLL polymers can be troublesome when both PNIPAm microgels and PLL chains are suspended in water ($n_{H2O}$ = 1.33), since PLL polymers have both a smaller size and a significantly lower refractive index contrast than microgels with respect to the solvent.

We now discuss more in detail the DLS results and report on the cumulant analysis of the correlograms for all the PLL-microgel mixtures. The unimodal size distributions and the limited polydispersity indexes (PDI < 0.4) allowed us to adopt this kind of analysis for all the samples.

\begin{figure}[htbp]
\centering
\includegraphics[width=12 cm]{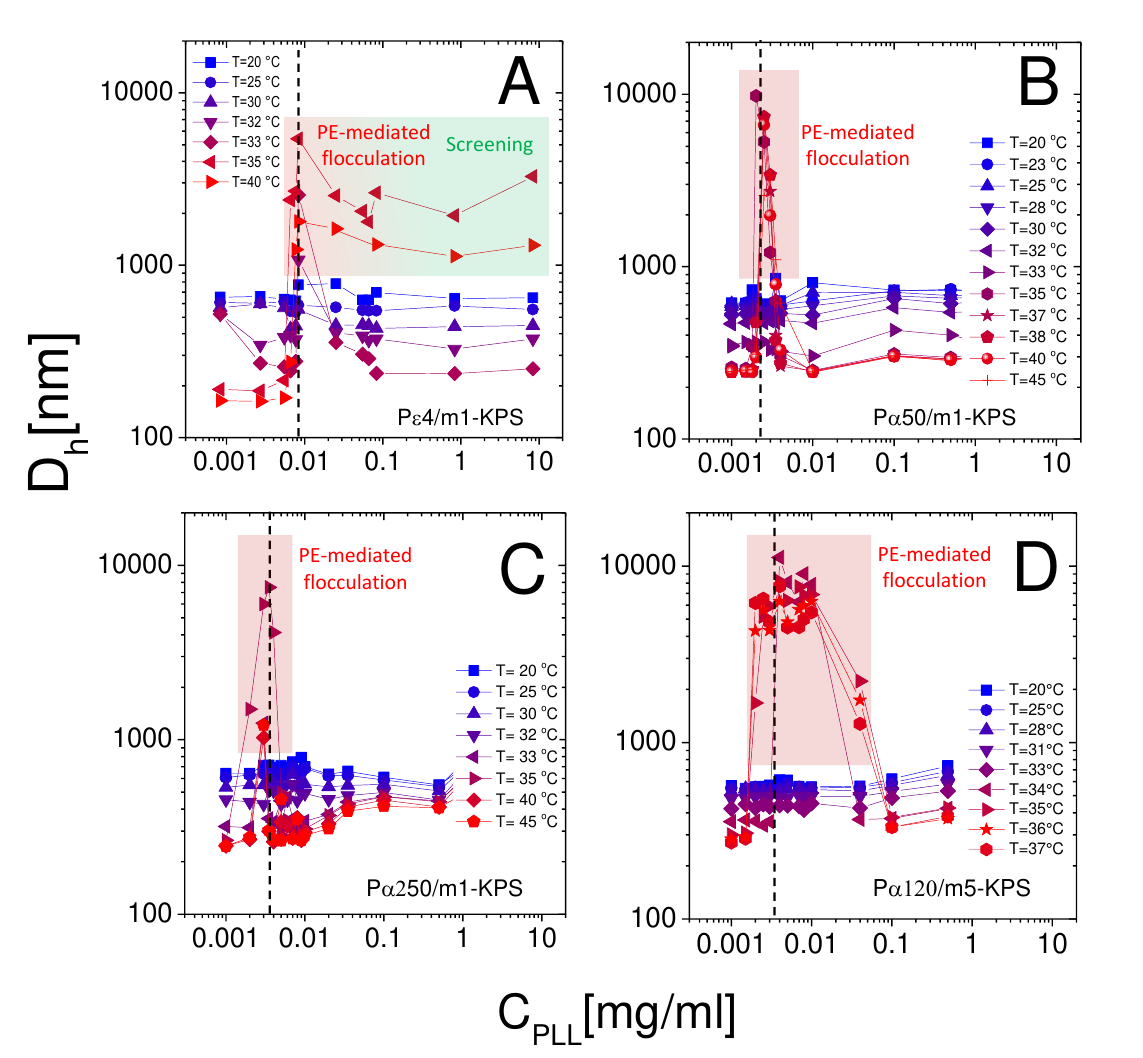}
\caption{Hydrodynamic diameters in function of $C_{PLL}$ for P$\varepsilon$4/m1-KPS (Panel \textbf{A}), P$\alpha$50/m1-KPS (Panel \textbf{B}), P$\alpha$250/m1-KPS (Panel \textbf{C}) and P$\alpha$120/m5-KPS (Panel \textbf{D}) mixtures, and selected temperatures as indicated in the panels. Points in the shaded regions refer to samples in which clustering at $T>T_c$ brought to the formation of macroscopic flocs. Vertical dashed lines corresponds to $C_{IEP}(40~^{\circ}$C). Data in panel A are adapted from~\cite{truzzolillo_overcharging_2018}-Published by The Royal Society of Chemistry (RSC) on behalf of the Centre National de la Recherche Scientifique (CNRS) and the RSC.}\label{sizeall}
\end{figure}

Figure~\ref{sizeall} shows the hydrodynamic diameter of PLL-microgel complexes as a function of PLL concentration for different temperatures. Although the average mobility reverses its sign at any temperature, only for $T \gtrsim T_c$ we observe reentrant condensation: complexes are stable for low PLL concentrations, they form large clusters or even unstable flocs where mobility is close to zero, and they re-stabilize for high PLL contents, where a large overcharging occurs. The presence of unstable flocs detected for P$\varepsilon$4/m1-KPS mixtures and high PLL concentrations has been already discussed in~\cite{truzzolillo_overcharging_2018} and is due to the residual free P$\varepsilon$4 chains that screen electrostatic repulsions between hydrophobic collapsed decorated microgels.

More specifically, at low temperatures ($T<T_c$) the addition of oppositely charged polymers does not produce flocculated microgels, the mobility is low in a wide range of polyelectrolyte concentrations and, despite a weak overcharging and an isolectric point can be identified, particle hydrophilicity is enough to keep the suspensions stable. 
Conversely, at high temperatures ($T>T_c$), the reentrant condensation of microgels signals the electrostatic nature of the assembly process~\cite{bordi_polyelectrolyte-induced_2009}. Such a finding validate a scenario where correlated adsorption takes place strictly above $T_c$, being enhanced by the collapse of microgels, and triggers the onset of charge patch attractions. Together with the aforementioned mobility behavior, this fully brings to light the double-faced electrostatic behavior of PNIPAm microgels in aqueous media.

In addition to that, our measurements allow debating the role played by the molecular weight of PLL chains in microgel reentrant clustering and charge inversion as well. As already mentioned, for the lowest PLL molecular weight (P$\varepsilon$4) the re-stabilization of microgels is not attained at large polymer contents and for the highest temperatures (\mbox{$T>33$ $^{\circ}$C}). This aspect can now be better analyzed in the light of the results obtained for the higher molecular weights. In fact, due to their low adsorption energy onto the microgel substrates, only a fraction of the dispersed P$\varepsilon$4 chains are adsorbed, leaving part of the collapsed microgel surfaces free with many chains remaining in the bulk, thus enhancing both charge heterogeneity and screening at high concentrations, eventually favoring aggregation.
As a matter of fact, in this case many free polyelectrolytes act as screening multivalent ions, giving rise to the same phenomenology observed at large NaCl molarity~\cite{truzzolillo_overcharging_2018}. The reduced extent of mobility reversal with respect to longer $\alpha$-polylysines (Figure~\ref{mobilityall}) is in line with the aforementioned limited adsorption and a concomitant intensified screening.
Such enhanced clustering at high $C_{PLL}$ is absent for larger molecular weight PLL polymers (Figure~\ref{sizeall}B--D), pointing to a more effective adsorption of longer chains.

Beyond that, we note that reentrant microgel aggregation is encountered also for P$\alpha$120/m5-KPS, with features analogous to P$\alpha$50/m1-KPS and P$\alpha$250/m1-KPS mixtures, suggesting once again the very general character of the mechanism driving the entire phenomenology. Nevertheless, in this case the extent of the instability region is seemingly larger with respect the other mixtures. Whether this is due the peculiar structure of m5-KPS microgels or to the existence of an \emph{optimum} molecular weight enhancing electrostatic patch attractions goes beyond the scope of this work, 
and it will not be investigated further, since here we are rather interested into the general features of the thermoresponsive electrostatic adsorption of PE chains onto microgels.

The differences characterizing the employed polymers, the occurrence of a crossover between poorly-adsorbed and highly-adsorbed PE chains and the impact of PLL molecular weight on the mobility reversal, can be highlighted even more if we consider the behavior of the isoelectric point (where $\mu=0$) for the different samples.

Figure~\ref{IEP} shows the PLL concentration $C_{IEP}$, where mobility reversal occurs, extrapolated by fitting linearly the two consecutive points with opposite signs around $\mu=0$ $\mu$mcm/Vs. For all the polylysine polymers employed here, $C_{IEP}$ drops in concomitance with the microgel collapse, showing that a much lower lysine content is needed to neutralize microgels when they are in their collapsed state. As a matter of fact, above $T_c$, due to the increased charged densities of microgels, we expect an increased capacity to adsorb stably oppositely charged macroions and, at the same time, a much larger correlation between the adsorbed chains. {Both such effects contribute to lower $C_{IEP}$~\cite{stout_continuum_2014}.} 

In this regard, it's worth pointing out that when microgels are brought above $T_c$ their size shrinks by a factor $\sim$2 giving rise to a charge density ranging from $\sim$4 to $\sim$8 times larger than the one at $T<T_c$, depending whether the charged initiators are distributed uniformly in the microgel volume or they are rather localized at the surface. As evident from the data reported in Figure~\ref{IEP}, there is a large $C_{IEP}$ difference between P$\varepsilon$4 and the other polymers at all temperatures. This finding can be rationalized considering the large difference in size and in valence  between the  short P$\varepsilon$4 and the other PLL samples.

As already argued in~\cite{truzzolillo_overcharging_2018}, the few ionized groups borne by short P$\varepsilon$4 chains are compatible with an adsorption energy of few $k_BT$ for $T<T_{c\mu}$.

Here we do a step further and estimate the critical molecular weight $M_w^c$ above which the adsorption energy is considerably larger than $k_BT$, namely when more than one ionized group on PLL chains can form an ion pair with the sulfate groups present in the microgels.

First of all, we remind that an approximate upper bound value for the charge-to-charge distance within the microgel can be computed supposing that the sulfate groups are distributed within all the microgel volume. {From the synthesis protocol~\cite{truzzolillo_overcharging_2018}, we know that each microgel bears a number of sulfate groups $Z$ ranging from $3.75 \times 10^5$ to $7.5 \times 10^5$ depending on the fraction of persulfate (or 2,2$'$-Azobis(2-methylpropionamidine)) molecules dissociating in solution during the synthesis. This gives an average charge-to-charge distance $d_{cc}=(\pi D_h^3/6Z)^{1/3}$ in the range 5.4--6.7 nm for $T=$25~$^{\circ}$C $<$ $T_{\mu c}$ and 2.2--2.7~nm for $T=$ 40~$^{\circ}$C $>$ $T_{\mu c}$.} Such distances must be compared with the size of a PLL chain that we consider equal to $2R_g$, with $R_g$ being the radius of gyration of the PLL polymers.
Moreover, we must note that the size of polyelectrolytes changes depending on whether they are free in bulk or adsorbed onto an oppositely charged surface~\cite{dobrynin_adsorption_2001}, with a size that depends on the charge density of the adsorbing substrate. This is not easily measurable or inferable for our soft microgels.
For such a reason we adopt another approximation and we consider $R_g$ of the chains as the one in the bulk. This is a lower bound for the real size of the adsorbed chains, since the latter are typically stretched when they lay on the adsorbing substrate. Both assumptions concerning microgel charge density and PLL size clearly lead to overestimate $M_w^c$ but allow computing its order of magnitude.
In~\cite{truzzolillo_overcharging_2018} we have already shown that according to the calculation based on these approximations, the charge density variation induced by VPT may bring a single P$\varepsilon$4 chain to neutralize more than one sulfate group anchored to the niPam network for $T>T_c$, thus reducing significantly the amount of chains needed to neutralize the whole microgel. Considering polylysine polymers as worm-like chains~\cite{jin_investigating_2014}, we can estimate the gyration radius as
\begin{equation}\label{rg}
    R_g=2l_p(N_k/6)^{1/2},
\end{equation}
where $l_p$ = 1.8 nm is the known persistence length measured for $\alpha$-polylysine chains~\cite{brant_configuration_1965} and $N_k$ is the number of statistical segments of the chains. $N_k$ is given by the ratio between the contour length of the polymer $L$ and the Kuhn length $l_k\simeq2l_p$ and reads:
\begin{equation}\label{Nk}
    N_k=\frac{L}{2l_p}=\frac{M_w\sigma_m}{2l_pM_w^m}
\end{equation}
where $M_w$ is the polyelectrolyte molecular weight, $M_w^m=146$ Da is the monomer molecular weight and $\sigma_m$ is the monomer diameter that has been estimated as the sum of the atomic covalent radii, which gives $\approx$0.6 nm. The critical molecular weight above which 2 ion pairs are formed is given by
\begin{equation}\label{Mwc}
    M_w^c=\frac{6d_{cc}^2M_w^m}{8l_p\sigma_m}.
\end{equation}

{Equation~(\ref{Mwc}) has been obtained by equating the chain size $2R_g$ to the maximum charge-to-charge distance on the microgel $d_{cc}$.
We obtain 2900 Da  $<M_w^c<$ 4510 Da for swollen microgels ($T<T_{c\mu}$) and 486 Da $<M_w^c<$ 732 Da for fully shrunken microgels ($T>T_{c\mu}$).}
{Although this is just a rough estimation for the critical molecular weight above which polyelectrolyte adsorption is expected to saturate, it captures reasonably well the behavior observed for PLL-microgel mixtures: all polylysine polymers with a molecular weight far larger than $M_w^c$, (namely P$\alpha$50, P$\alpha$120 and P$\alpha$250) neutralize microgels at essentially the same $C_{IEP}$ for all temperatures, while a much larger amount of lysines monomers are needed to attain microgel neutralization if they are added in the form of P$\varepsilon$4 chains, the latter having a molecular weight comparable to or lower than $M_w^c$, and hence an adsorption energy comparable to $k_BT$, especially for $T<T_c$.}

More specifically, P$\varepsilon$4 chains bring microgels to zero mobility at concentrations about one order of magnitude larger than those measured for the longer polymers below $T_c$, such a gap becoming narrower for $T>T_c$. 
{Our estimate therefore suggests that the ratio between the polyelectrolyte size and the inter-charge distance on the microgels is a crucial parameter that must be taken into consideration in predictive models for the adsorption of charged polymers on microgel networks, and points further to an electrostatic adsorption onto microgel particles driven by a fine thermodynamic balance between the adsorbed and the free (bulk) state of the chains. On that note, a more detailed knowledge of the charge distribution within the microgels would surely improve any prediction regarding polyelectrolyte adsorption.} 
Finally, we may note that in the range 35 $^{\circ}$C $<$ $T$ $<$ 45~$^{\circ}$C and for all the $\alpha$-PLL polymers, $C_{IEP}$ weakly increases as a function of temperature. We attribute this effect to three synergic phenomena. (i) Since microgels continue, albeit weakly, to shrink for $T>T_c$, the reduced space available for the chains on each microgel peripheral surface may impact the amount of chains that one microgel can adsorb, while inter-chains repulsions, both steric and electrostatic, hamper a further adsorption. (ii) Collapsed microgels become also less penetrable and the chains have on average less access to the oppositely charged groups borne by the microgels periphery. This may cause a weak decrease of the (negative) charge available for complexation. (iii) The adsorption of hydrophilic polymers, as PLLs, on less hydrophilic substrates is less energetically favored than on swollen and more hydrophilic ones.

\begin{figure}[htbp]
\centering
 \includegraphics[width=15 cm]{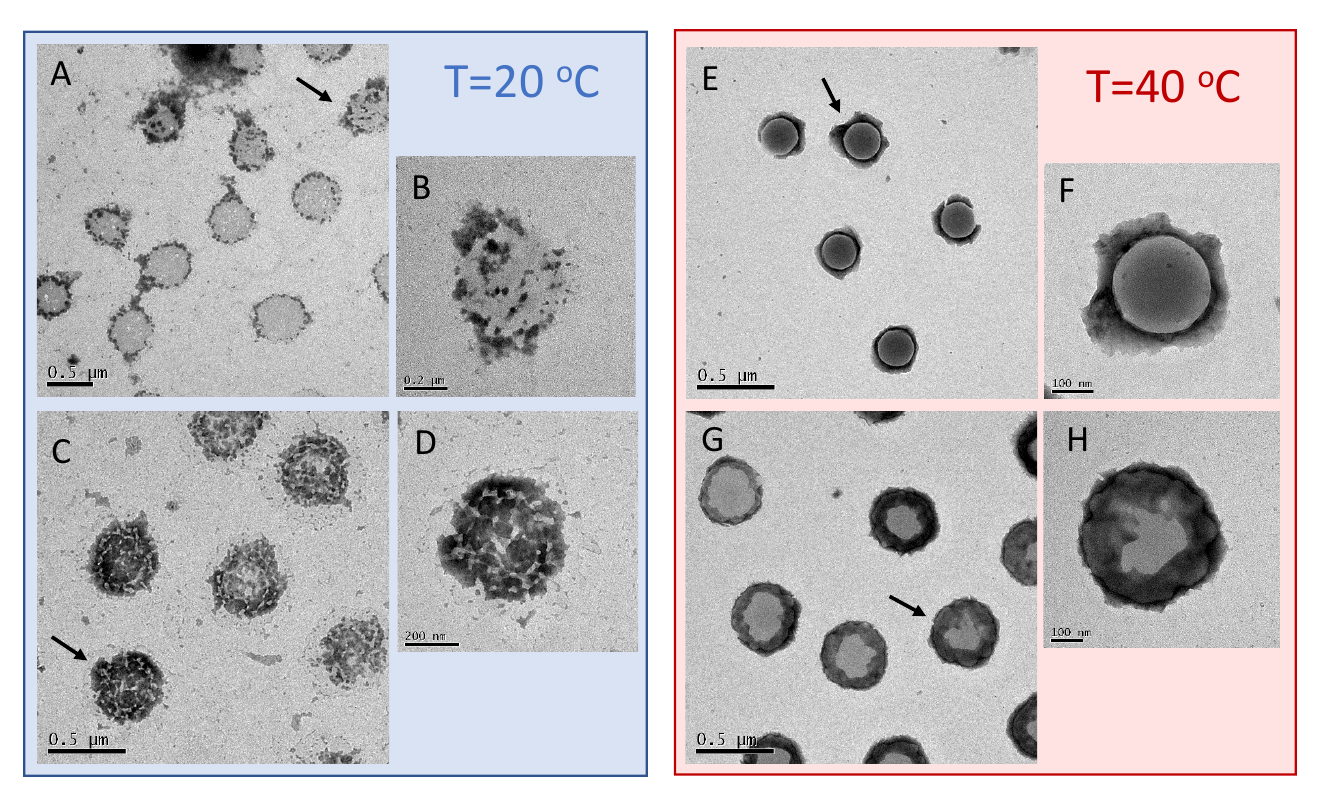}\\
\caption{TEM images (PTA staining) of P$\alpha$250/m1-KPS complexes formed at C = 0.0015 mg/mL (panels \textbf{A},\textbf{B} and \textbf{E},\textbf{F}) and C = 0.01 mg/mL (panels \textbf{C},\textbf{D} and \textbf{G},\textbf{H}), for $T<T_c$ (left block) and $T>T_c$ (right block). The interaction between PTA and PLL chains originates the positive staining of polycation-covered microgel and regions covered by P$\alpha$250 chains appear as dark halo surrounding or overlaying particles. At $T<T_c$ these regions assume an evident patchy-like pattern, while above $T_c$ they become more homogeneously distributed around the outer shell of the microgels.}\label{TEM}
\end{figure}

\begin{figure}[htbp]
\centering
 \includegraphics[width=15 cm]{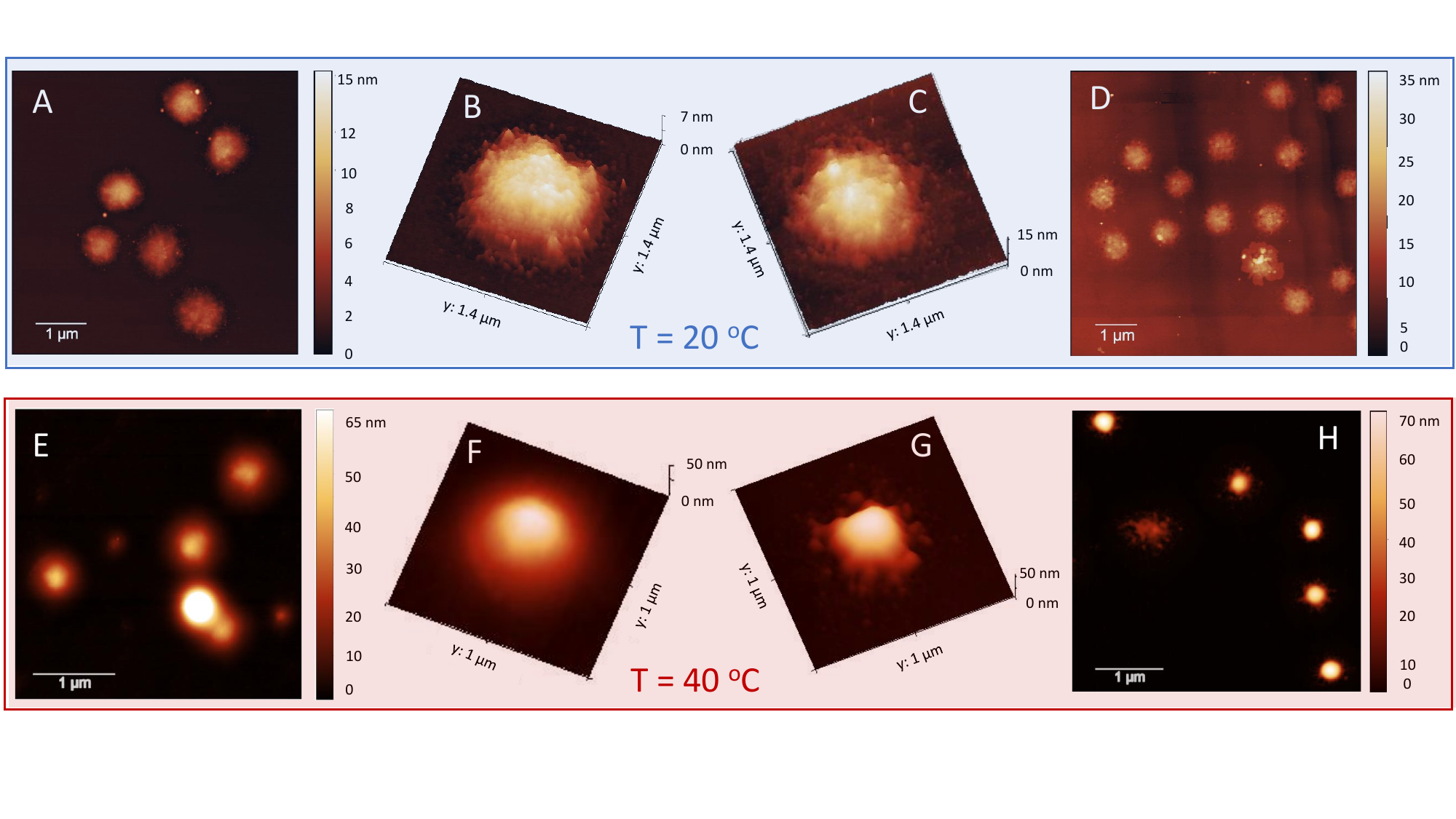}
\caption{AFM images of P$\alpha$250/m1-KPS complexes formed at  C = 0.0015 mg/mL (panels \textbf{A},\textbf{B} and \textbf{E},\textbf{F}) and \mbox{C = 0.01 mg/mL} (panels \textbf{C},\textbf{D} and \textbf{G},\textbf{H}), for $T<T_c$ (upper panel) and $T>T_c$ (lower panel). The presence of the PLL shell surrounding microgels is particularly evident above $T_c$ where a diffuse or globular-like corona is observed depending on PLL amount.}\label{AFM}
\end{figure}

To characterize even further and visualize the adsorption of long PLL chains on the microgels, we performed TEM and AFM imaging. To this aim, we deposited on a TEM grid and a mica support an aliquot of P$\alpha$250/m1-KPS suspensions at two concentrations, $C_{PLL}=$ 0.0015 mg/mL and $C_{PLL}=$ 0.01 mg/mL, i.e., below and above the condensation region respectively, and we performed TEM and AFM imaging. The depositions and the successive drying have been performed at 25 $^\circ$C and 40 $^\circ$C, as described in Section~\ref{Mic}. Representative images are displayed in Figures~\ref{TEM} and~\ref{AFM}.
In TEM images, due to PTA staining, microgel particles appear light grey while the positively
charged P$\alpha$250 chains appear as darker knots since they are able to attract the negatively charged PTA, as we have shown by ESI spectroscopy  in our previous investigation on P$\varepsilon$4~\cite{truzzolillo_overcharging_2018}. The differences between the two selected concentrations and the two temperatures appear evident: by increasing the PLL concentration and crossing the condensation region more chains are adsorbed onto m1-KPS microgels, with a PLL distribution passing from a sparse and patchy decoration below $T_c$ (Figure~\ref{TEM}A--D) to a more homogeneous coverage above $T_c$ (Figure~\ref{TEM}E--H). Here a more pronounced coffee-stain effect is also present, since microgels are both more hydrophobic and less spread on the substrate. AFM images (Figure~\ref{AFM}) confirm the variation of surface morphology of P$\alpha$250/m1-KPS complexes,  as it can be seen by the large view images showing several particles, and by the 3D AFM maps of single complexes. As already observed for bare microgels, below $T_c$ the affinity with the mica surface induces an evident flattening of the swollen P$\alpha$250/m1-KPS particles.  Upon collapse, the net contrast between the hydrophobic microgel and the hydrophilic PLL chains, sliding down during the drying process, gives rise to different sombrero-like morphologies in dependence on PLL concentration, due to a different water retention in the external shell of the P$\alpha$250/m1-KPS complexes. It is also interesting to note that in all cases PLL polymers mainly accumulate around the microgels and very few chains are free,  as it is observed by the almost flat appearance of mica support.
This is in contrast with what we have reported in~\cite{truzzolillo_overcharging_2018} for P$\varepsilon$4 polymers, for which the presence of free chains has been detected below $T_c$, corroborating once again the scenario in which the molecular weight of the polyelectrolytes tunes the their adsorption energy and regulates the balance between the adsorbed and the free state of the chains.


\nointerlineskip


\nointerlineskip
\vspace{-12pt}

\subsection{Nanoparticle-Microgel Suspensions}\label{microLudox}
\begin{figure}[htbp]
  \centering
 \includegraphics[width=15 cm]{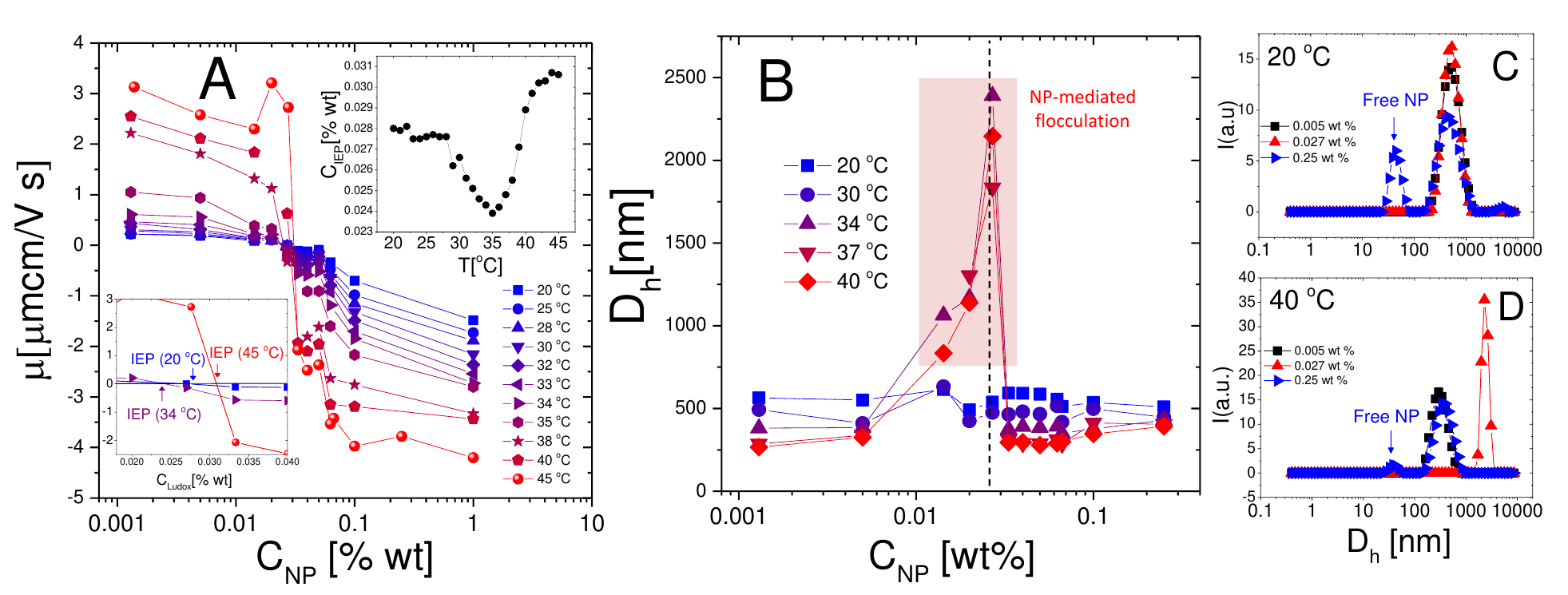}
\caption{Electrophoretic mobility (Panel \textbf{A}) and hydrodynamic diameter (Panel \textbf{B}) of microgels in function of the co-suspended Ludox weight fraction for different temperatures as indicated in the panels. The insets in panel A show the concentration $C_{IEP}$ where electrophoretic mobility is zero as a function of temperature (top inset) and a zoom of the isoelectric region for three representative temperatures (bottom inset).
Symbols within the shaded region in panel B refer to samples in which clustering at $T>T_c$ brought to the formation of sedimenting flocs. The vertical line corresponds to $C_{IEP}(40~^{\circ}$C).
Panels \textbf{C} and \textbf{D} show the intensity-weighted distribution obtained via the CONTIN analysis of the intensity correlation function $g_2(\tau)$-1 at 20 $^{\circ}$C and 40 $^{\circ}$C and three NP concentrations correspondent to poorly NP-decorated microgels, nearly neutralized microgels and overcharged microgels. {As in PLL-microgel mixtures, NP-microgel complexation is characterized by both reentrant aggregation and charge reversal, however, in contrast to polyelectrolyte-microgel complexes, not all nanoparticles are adsorbed, even at $T>T_c$, with a reduced mobility change.}}\label{microludox}

\end{figure}

To probe the general character of the phenomenology just described, anionic Ludox TM nanoparticles have been added to suspensions of cationic m5-AIBA microgels. \mbox{Figure~\ref{microludox}} shows the average electrophoretic mobility (Panel A), the measured hydrodynamic diameter (Panel B) for selected NP-m5-AIBA samples in function of NP weight fraction and different temperatures and the intensity-weighted size distributions for three samples below (Panel C) and above (Panel D) $T_c$. The hydrodynamic diameter of NP-microgel complexes has been obtained through the CONTIN analysis of the correlograms since part of the dispersed NP stayed unadsorbed and contribute to the light scattered by the sample, especially at high NP concentrations (Figure~\ref{microludox}C,D) and $T<T_c$. By contrast, since the contribution of microgels or NP-microgel complexes to the scattered intensity dominates over that of free NPs due to their two radically different size distributions, PALS measurements~\cite{varenne_evaluation_2019} do not allow to resolve free NP particle mobility in the mixtures. This technique indeed, allows discerning different electrophoretic mobility populations mostly for similarly sized particles characterized by comparable number density and radically different charge~\cite{liu_bitumenclay_2002,liang_heterocoagulation_2017}.
The encountered phenomenology is therefore very similar to that found in PLL-microgel mixtures, but with reversed mobilities:
m5-AIBA microgels are stable at low temperatures ($T<T_c$) and the electrophoretic mobility shows a weak, albeit present, dependence on NP concentration (Figure~\ref{microludox}A,B). At higher temperatures ($T>T_c$) a sharp change of the electrophoretic mobility appears together with a region of instability, confirming that in the presence of large macroion correlation on top of a highly charged adsorption surface (the peripheral networks of microgels here), charge patch attraction destabilizes colloidal suspensions and a large colloidal overcharging occurs. Despite these similarities and a mobility behavior consistent with a large variation of the spatial correlation of NPs, we note a quite striking difference between PLL-microgel and NP-microgel mixtures described herein. The variation of $C_{IEP}$, where mobility inversion occurs, is not as large as in PLL-microgel mixtures: while $C_{IEP}$ drops by a factor ranging from $\approx$10 to $\approx$2, respectively for P$\varepsilon$4/m1-KPS and P$\alpha$120/m5-KPS mixtures, this drop is much less pronounced for NP/m5-AIBA mixtures (insets of Figure~\ref{microludox}A) for which $C_{IEP}$ shows a clear minimum close to the electrokinetic transition of the microgels. Such a finding suggests that nanoparticles neutralize and adsorb onto PNIPAm microgels to a large extent also at low temperatures, microgel VPT does not affect much the amount of particles needed to counterbalance the charge of microgels and that further (and massive) NP adsorption is somehow hampered at $T>T_c$.

More in detail, due to this weak drop of $C_{IEP}$ and a presumable more efficient steric and electrostatic repulsion between silica colloids, the relative increase of $C_{IEP}$ above $T_{c\mu}$ is more pronounced than in PLL-microgel systems and, at the same time confirms the result already obtained for the longest polylysine chains: collapsed microgels with their compact surface do not favor the adsorption of oppositely charged species if the latter have enough steric and electrostatic hindrance and presumably less access to the microgel charges. The presence of free silica particles both below and above $T_c$ for high enough NP concentrations (Figure~\ref{microludox}C,D), where microgels show reversed mobility, supports such a scenario.

\begin{figure}[htbp]
  \centering
 \includegraphics[width=14 cm]{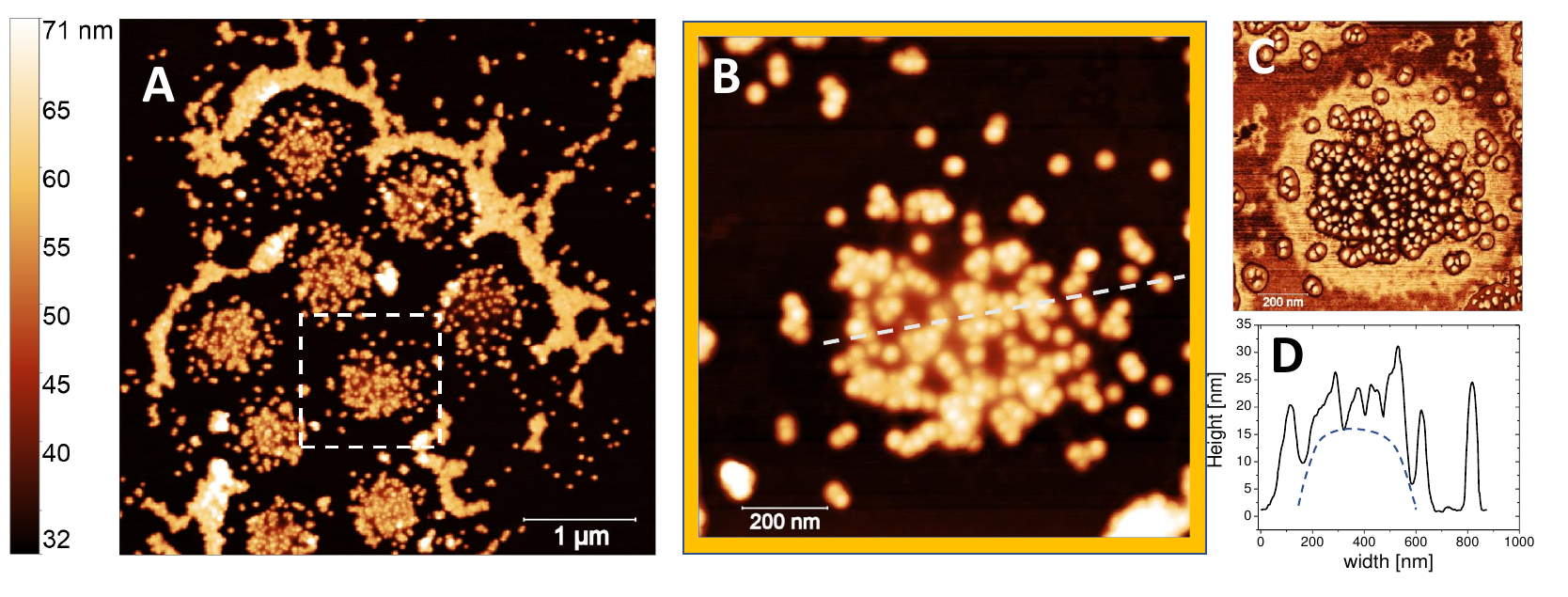}
\caption{AFM images of NP-microgel complexes formed at $T<T_c$ at 0.1\% NP weight fraction. Upon drying several NP-microgel complexes appear arranged in an ordered configuration surrounded by free and close-packed NP clusters (panel \textbf{A}). Panel \textbf{B} shows a high magnification of the white-squared region in panel \textbf{A}. The phase image and the height profile of a single NP-microgel complex along the line shown in panel B are displayed in panels \textbf{C} and \textbf{D} respectively. In panel \textbf{D}, the dashed line following a spheroidal shape is a guide for eye, only.}\label{AFMludox}
\end{figure}

To visualize NP/m5-AIBA complexes and support our finding in bulk we performed AFM imaging after having dried at T = 25 $^\circ$C an aliquot of a sample at 0.1\% NP weight fraction, that is a concentration well above $C_{IEP}$ (see Figure~\ref{microludox}). The drying process causes the accumulation of both NP/m5-AIBA microgels and free NP (Figure~\ref{AFMludox}A). Notably, the formation of an ordered configuration where several NP/m5-AIBA microgels stay close to each other without sticking together is in line with the presence of repulsive interactions. Conversely, free silica nanoparticles accumulate in a distinct drying pattern formed by several arrays of closely-packed particles. The AFM image of an individual NP/m5-AIBA complex is displayed in panel B of Figure~\ref{AFMludox}, and clearly shows the non-homogeneous decoration  by the adsorbed NP. By inspecting the corresponding phase image (Figure~\ref{AFMludox}C), thanks to the different interaction of the AFM tip with the microgels and mica surface, it is possible to evidence that distinct patches of adsorbed NP can be found only within a circular region delimited by the external shell of microgel particle. The occurrence of NP adsorption onto cationic microgels is also supported by the height profile (Figure~\ref{AFMludox}D) traced along one diameter of a NP/m5-AIBA complex, where small height bumps of about 20 nm stand out against the spheroidal dried microgel.

This said, a more detailed investigation of the structure of NP-microgel complexes is required to achieve a deeper understanding of NP adsorption on soft PNIPAm microgels and we reserve the right to report further results on these systems in a future study.

\nointerlineskip

\subsection{Thermal Reversibility of Overcharged Complexes}\label{reversibility}
\begin{figure}[htbp]
\centering
 \includegraphics[width=16 cm]{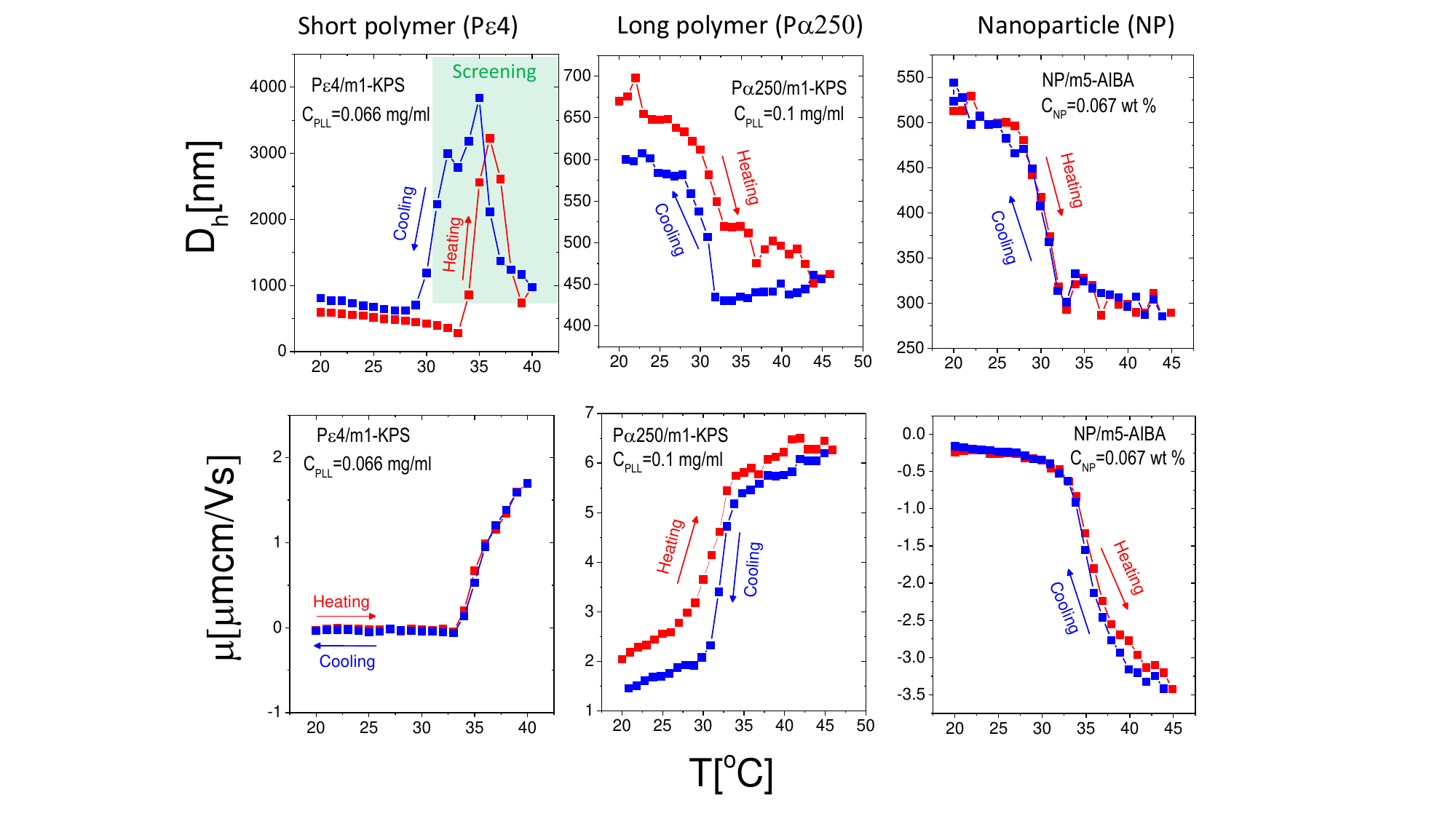}
\caption{Hydrodynamic diameter $D_h$ (\textbf{upper panels}) and electrophoretic mobility $\mu$ (\textbf{lower panels}) as a function of temperature during
heating (red squares) and cooling (blue squares) ramps with thermalization time 300 s for P$\varepsilon$4/m1-KPS (\textbf{left panels}), P$\alpha$250/m1-KPS (\textbf{middle panels}) and NP/m5-AIBA (\textbf{right panels}). Additive concentrations are indicated in the panels and correspond to the formation of overcharged complexes. Symbols in the shaded region for P$\varepsilon$4/m1-KPS mixtures correspond to samples where flocculation occurred because of the high screening given by the free P$\varepsilon$4 chains. {A pronounced hysteretic behavior of both mobility and size has been detected only for P$\alpha$250/m1-KPS mixtures, suggesting different time scales of adsorption and desorption of long polyelectrolytes. By contrast, the full reversibility obtained for NP/m5-AIBA complexes points to a quasi-instantaneous rearrangement of NP as temperature is increased or decreased.}}\label{reverso}
\end{figure}

To test the thermal reversibility of the complexation between microgels and additives (PE and NP), we have performed temperature cycles
for selected concentrations of additives giving rise to overcharged complexes for P$\varepsilon$4/m1-KPS (data adapted from~\cite{truzzolillo_overcharging_2018}), P$\alpha$250/m1-KPS and NP/m5-AIBA systems, being the mixtures containing the shortest polymer, the longest polymer and the nanoparticles investigated in this work respectively.
Thermal cycles were performed according to the following protocol: (i) a first ascending ramp from 20 $^{\circ}$C to 45 $^{\circ}$C
by increasing the temperature by 1 $^{\circ}$C each time. Before each measurement the samples were left to thermalize for 300 s at
the target temperature (the standard protocol already described in Section~\ref{dls}); (ii) a descending ramp from 45 $^{\circ}$C down to 20 $^{\circ}$C followed, with the same temperature step and thermalization time of (i). The results are shown in Figure~\ref{reverso}.

\nointerlineskip


First, we note that, as expected from the results reported in Section~\ref{microPLL}, for P$\varepsilon$4/m1-KPS mixtures overcharged complexes flocculate as they are heated up above $T_c$. However, although the original microgel size is not fully recovered within the time of our experiment, microgel condensation is reversible as shown by the decrease of the hydrodynamic diameter below 1 $\mu$m for $T<T_c$. This uncomplete clustering dissolution is presumably due to the different time scales characterizing the adsorption of the polyelectrolyte on the external shell of the microgels and full cluster dissolution. The first is driven by both polyion and single microgel diffusion and gives rise to the (almost instantaneous) aggregation of decorated microgels as the temperature is raised above $T_c$. The rapid cluster formation as polyelectrolytes are mixed with oppositely charged
colloids, has been observed in all polyelectrolyte-colloid mixtures and has been discussed within the framework of kinetically
arrested (metastable) clustering~\cite{bordi_polyelectrolyte-induced_2009}. By contrast, cluster dissolution is driven by both the time scale associated with polymer desorption from the microgel and the one associated with microgel intra-cluster diffusion after cooling the system. The latter, being dominated by the complete disentanglement of aggregated microgels, is a much larger time scale than the former. In addition to that, mobility shows full reversibility, pointing to a rapid adsorption/desorption of P$\varepsilon$4 chains during the cycle. This aspect has been discussed more widely in~\cite{truzzolillo_overcharging_2018}.
For larger polymers, where stable PE-microgel complexes form (Figure~\ref{reverso}, middle panels), an hysteretic behavior of both size and mobility characterizes the heating-cooling cycle: after the first (ascending) ramp, upon ramp cooling down the samples the size and the mobility of the complexes differ from the values measured when temperature is increased. We note in particular that P$\alpha$250-decorated microgels stay slightly more shrunken during the cooling-down process and their mobility is slightly lower than during heating up at basically all temperatures.
We attribute this behavior to the fact that P$\alpha$250 chains, adsorbed at high temperature, do not allow the instantaneous re-swelling
of the microgels, that, for such a reason, stay more densely charged and give rise to lower mobility values. We remind here that m1-KPS microgels are negatively charged, and hence a reduction of their size produces a decrease of the overall mobility of the PE-microgel complex. This conforms to a non-instantaneous microgel re-swelling where only the most peripheral adsorbed PLL chains detach rapidly, leaving microgels temporarily contracted. 
Finally, NP/m5-AIBA complexes show full reversibility: both mobility and size return to their initial values once the suspensions are cooled down after the first ascending temperature ramp. However, while the size of the complexes does not show hysteresis, we note that the mobility shows a very weak hysteretic behavior for $T>T_c$, where $\mu$ measured during the cooling process is slightly larger in magnitude (at most 15\% of relative variation) than during the ascending ramp. Yet, since the mobility difference between the heating and the cooling ramps is small, our results are far from being conclusive and do not allow speculating on the different adsorption/desorption kinetics in this case, that would require a more extensive investigation including the adoption of different heating-cooling protocols. This said, by comparing the different systems, we can state that adsorption and desorption of NP, and the consequent rearrangement of the overall charge density in NP/m5-AIBA decorated microgels are faster processes as compared to the formation and the dissolution of P$\alpha$250/m1-KPS complexes: nanoparticles desorb and/or rearrange more quickly than long polyelectrolytes on ionic microgel surfaces.

\section{Conclusions}\label{conclusions}
In this work we have reported on the electrostatic complexation of ionic PNIPAm microgels with oppositely charged polyelectrolytes and nanoparticles. By systematically monitoring the cluster formation and the electrophoretic behavior in PE- and NP-microgel complexes by means of dynamic light scattering, electrophoresis, transmission electron microscopy and atomic force microscopy experiments, we have discerned two distinct regimes of complexation for PNIPAm microgels synthesized by free radical polymerization. On the one hand, we have shown that when microgels are swollen and polyelectrolytes or nanoparticles are progressively added, the measured electrophoretic mobility is subject to a weak reversal and clusters never form. On the other hand when microgels are in their collapsed state, they strongly interact with oppositely charged PE or NP, a large and sharp mobility inversion occurs and clusters form only close to the isoelectric point, i.e., where the mobility of PE-microgel or NP-microgel complexes is zero: A strong overcharging and a colloidal reentrant condensation are encountered simultaneously as in suspensions of densely charged colloids.
By employing polylysine polymers of different molecular weight, we have further pointed out that charge compensation and colloidal condensation are essentially unaltered by varying polyelectrolyte length above a certain threshold size of the added polymers. The scenario that we propose includes a thermosensitive electrostatic self-assembly driven by the correlated adsorption of PE-chains or NP onto soft PNIPAm microgels, where adsorption energy is dictated both by the discrete charge distribution of the microgels and that of the adsorbing species.
{By paving the way to the study of reversible electrostatic complexation between PNIPAm microgels and oppositely charged species, our work opens many complementary lines of research. In fact there are many aspects of stimuli-sensitive electrostatic adsorption that can be investigated in the near future and that concern both fundamental and applicative purposes. Among them, we would like to cite (i) the study of the role of microgel softness in determining the extent and the reversibility of polyelectrolyte and nanoparticle adsorption, (ii) the role played by charge distribution, namely the role of the fraction of ionizable co-monomers that can be embedded within PNIPAm networks on PE and NP adsorption, and (iii) the effect of adsorbing charged agents on the microscopic dynamics and rheology of glassy microgel suspensions.
Motivated by all these perspectives, we hope that our work can inspire further efforts to investigate the tunable electrostatic properties of PNIPAm microgels.}

\vspace{6pt}



\textbf{Author contributions}: S.S. and D.T. designed and performed the DLS and electrophoresis experiments; E.C. performed microgel synthesis; S.C. and S.S. performed TEM and AFM measurements, respectively. D.T. wrote the original draft; S.S. and F.B. reviewed and edited the manuscript. All authors have read and agreed to the published version of the manuscript.\\\\

\textbf{Funding}: This research has been funded by the Agence Nationale de la Recherche (ANR) (Grant N$^{\circ}$ ANR-20-CE06-0030-01, THELECTRA) and the European Research Council (ERC Consolidator Grant 681597, MIMIC).\\\\

\acknowledgments{D.T. warmly acknowledges \emph{Sapienza---University of Rome} for the visiting research fellowship granted from October 2018 to December 2018. S.S. thanks the \emph{Interdepartmental research center on nanotechnologies applied to engineering of Sapienza (C.N.I.S.)} for providing the AFM apparatus. D.T. and S.S. are grateful to Emanuela Zaccarelli for enlightening discussions and Lorenzo Rovigatti for graphic support.}

\end{document}